\documentstyle[sprocl,epsf,epsfig]{article}

\def\Journal#1#2#3#4{{#1} {\bf #2}, #3 (#4)}

\def\PLB{{\em Phys. Lett.}  B}

\def\PRD{{\em Phys. Rev.} D}

\def\ra{\rightarrow}

\def\be{\begin{equation}}
\def\ee{\end{equation}}
\def\bea{\begin{eqnarray}}
\def\eea{\end{eqnarray}}
\def\gappeq{\mathrel{\rlap {\raise.5ex\hbox{$>$}} {\lower.5ex\hbox{$\sim$}}}}
\def\lappeq{\mathrel{\rlap{\raise.5ex\hbox{$<$}} {\lower.5ex\hbox{$\sim$}}}}
\def\beq{\begin{equation}}
\def\eeq{\end{equation}}
\def\bea{\begin{eqnarray}}
\def\eea{\end{eqnarray}}
\def\bq{\begin{quote}}
\def\eq{\end{quote}}

\begin{document}

\title{THE STANDARD ELECTROWEAK THEORY AND BEYOND}

\author{G. ALTARELLI}

\address{Theoretical Physics Division, CERN\\
CH-1211 Geneva 23\\and\\
Universit\`a di Roma Tre, Rome, Italy \\E-mail: gual@mail.cern.ch} 

%\author{A. N. OTHER}

%\address{Department of Physics, Theoretical Physics, 1 Keble Road,\\
%Oxford OX1 3NP, England\\E-mail: gual@mail.cern.ch}


\maketitle\abstracts{ }
%%%%%%%%%%%%%%%%%%%%%%%%%%%%%%%%%%%%%%%%%%%%%%%%%%%%%%%%%%%%%%%%%%%%%%%%%%%%%%
\section{Introduction}
These lectures on electroweak (EW) interactions start with a short summary of the Glashow--Weinberg--Salam theory 
and then cover in detail some main subjects of present interest in phenomenology.

The modern EW theory inherits the phenomenological successes of the $(V-A) \otimes (V-A)$ four-fermion
low-energy description of weak interactions, and provides a well-defined and consistent theoretical framework
including weak interactions and quantum electrodynamics in a unified picture.

 As an introduction, we recall some salient physical features of the weak interactions. The weak
interactions derive their name from their intensity. At low energy the strength of the effective four-fermion
interaction of charged currents is determined by the Fermi coupling constant $G_F$. For example, the effective
interaction for muon decay is given by
\beq {\cal L}_{\rm eff} = (G_F/\sqrt 2) \left[ \bar
\nu_{\mu}\gamma_{\alpha}(1-\gamma_5)\mu \right]
\left[ \bar e\gamma^{\alpha}(1-\gamma_5)\nu_e \right]~,
\label{1}
\eeq with \cite{pdg} 
\beq G_F = 1.16639(1) \times 10^{-5}~{\rm GeV}^{-2}~.
\label{2}
\eeq In natural units $ \hbar = c = 1$, $G_F$ has dimensions of (mass)$^{-2}$. As a result, the intensity of weak
interactions at low energy is characterized by
$G_FE^2$, where $E$ is the energy scale for a given process ($E \approx m_{\mu}$  for muon decay). Since
\begin{equation} G_FE^2 = G_Fm^2_p(E/m_p)^2 \simeq 10^{-5}(E/m_p)^2~,
\label{3}
\end{equation} where $m_p$ is the proton mass, the weak interactions are indeed weak at low energies (energies of order
$m_p$). Effective four
fermion couplings for neutral current interactions have comparable intensity and energy behaviour. The quadratic increase with
energy cannot continue for ever, because it would lead to a violation of unitarity. In fact, at large energies the propagator
effects can no longer be neglected, and the current--current interaction is resolved into current--$W$ gauge boson vertices
connected by a $W$ propagator. The strength of the weak interactions at high energies is then measured by $g_W$, the
$W-\mu$--$\nu_{\mu}$ coupling, or, even better, by
$\alpha_W = g^2_W/4\pi$ analogous to the fine-structure constant $\alpha$ of QED. In the standard EW theory, we have
\begin{equation}
\alpha_W = \sqrt 2~G_F~\frac{m^2_W}{\pi} = \frac{\alpha}{\sin^2\theta_W} \cong 1/30~.
\label{4}
\end{equation} That is, at high energies the weak interactions are no longer so weak.

 The range $r_W$ of weak interactions is very short: it is only with the experimental discovery of the $W$ and $Z$
gauge bosons that it could be demonstrated that $r_W$ is non-vanishing. Now we know that
\begin{equation}
 r_W = \frac{\hbar}{m_Wc} \simeq 2.5 \times 10^{-16}~{\rm cm}~,
\label{5}
\end{equation} corresponding to $m_W \simeq 80$~GeV. This very large value for the $W$ (or the
$Z$) mass makes a drastic difference, compared with the massless photon and the infinite range of the QED force. The
direct experimental limit on the photon mass is \cite{pdg}
$m_{\gamma} <2~10^{-16}~eV$. Thus, on the one hand, there is very good evidence that the photon is massless. On the
other hand, the weak bosons are very heavy. A unified theory of EW interactions has to face this striking
difference.

Another apparent obstacle in the way of EW unification is the chiral structure of weak interactions: in the
massless limit for fermions, only left-handed quarks and leptons (and right-handed antiquarks and antileptons) are
coupled to $W$'s. This clearly implies parity and charge-conjugation violation in weak interactions.

The universality of weak interactions and the algebraic properties of the electromagnetic and weak currents [the
conservation of vector currents (CVC), the partial conservation of axial currents (PCAC), the algebra of currents,
etc.] have been crucial in pointing to a symmetric role of electromagnetism and weak interactions at a more fundamental
level. The old Cabibbo universality for the weak charged current:
\begin{eqnarray} J^{\rm weak}_{\alpha} &=&
\bar \nu_{\mu}\gamma_{\alpha} (1-\gamma_5)\mu +
\bar \nu_e\gamma_{\alpha}(1-\gamma_5) e +
\cos\theta_c~\bar u \gamma_{\alpha}(1-\gamma_5)d + \nonumber \\  &&\sin \theta_c~\bar u \gamma_{\alpha}(1-\gamma_5)s +
...~,
\label{6}
\end{eqnarray} suitably extended, is naturally implied by the standard EW theory. In this theory the weak
gauge bosons couple to all particles with couplings that are proportional to their weak charges, in the same way as the
photon couples to all particles in proportion to their electric charges [in Eq.~(\ref{6}), $d' =
\cos\theta_c~d + \sin \theta_c~s$ is the weak-isospin partner of $u$ in a doublet.  The $(u,d')$ doublet has the same
couplings as the $(\nu_e,\ell)$ and 
$(\nu_{\mu},\mu)$ doublets].

 Another crucial feature is that the charged weak interactions are the only known interactions that can change flavour:
charged leptons into neutrinos or up-type quarks into down-type quarks. On the contrary, there are no flavour-changing
neutral currents at tree level. This is a remarkable property of the weak neutral current, which is explained by the
introduction of the Glashow-Iliopoulos-Maiani mechanism and has led to the successful prediction of charm.

 The natural suppression of flavour-changing neutral currents, the separate conservation of $e, \mu$  and $\tau$
leptonic flavours, the mechanism of CP violation through the phase in the quark-mixing matrix, are all crucial
features of the Standard Model. Many examples of new physics tend to break the selection rules of the standard theory.
Thus the experimental study of rare flavour-changing transitions is an important window on possible new physics.

 In the following sections we shall see how these properties of weak interactions fit into the standard EW
theory.

\section{Gauge Theories}

In this section we summarize the definition and the structure of a gauge Yang--Mills theory. We
will list here the general rules for constructing such a theory. Then in the next section these results will be applied
to the EW theory.

Consider a Lagrangian density ${\cal L}[\phi,\partial_{\mu}\phi]$ which is invariant under a $D$ dimensional continuous
group of transformations:
\begin{equation}
\phi' = U(\theta^A)\phi\quad\quad (A = 1, 2, ..., D)~.
\label{7}
\end{equation} For $\theta^A$ infinitesimal, $U(\theta^A) = 1 + ig \sum_A~\theta^AT^A$, where
$T^A$ are the generators of the group $\Gamma$ of transformations in the (in general reducible)
representation of the fields $\phi$. Here we restrict ourselves to the case of internal symmetries, so that $T^A$ are
matrices that are independent of the space--time coordinates. The generators $T^A$ are normalized in such a way that
for the lowest dimensional non-trivial representation of the group $\Gamma$ (we use $t^A$ to denote the generators in
this particular representation) we have
\begin{equation} {\rm tr}(t^At^B) = \frac{1}{2} \delta^{AB}~.
\label{8}
\end{equation} The generators satisfy the commutation relations
\begin{equation} [T^A,T^B] = iC_{ABC}T^C~.
\label{9}
\end{equation} In the following, for each quantity $V^A$ we define
\begin{equation} {\bf V} = \sum_A~T^AV^A~.
\label{10}
\end{equation} If we now make the parameters $\theta^A$ depend on the space--time coordinates
$\theta^A = \theta^A(x_{\mu}),$ ${\cal L}[\phi,\partial_{\mu}\phi]$ is in general no longer invariant under the gauge
transformations $U[\theta^A(x_{\mu})]$, because of the derivative terms. Gauge invariance is recovered if the ordinary
derivative is replaced by the covariant derivative:
\begin{equation} D_{\mu} = \partial_{\mu} + ig{\bf V}_{\mu}~,
\label{11}
\end{equation} where $V^A_{\mu}$ are a set of $D$ gauge fields (in one-to-one correspondence with the group generators)
with the transformation law
\begin{equation} {\bf V}'_{\mu} = U{\bf V}_{\mu}U^{-1} - (1/ig)(\partial_{\mu}U)U^{-1}~.
\label{12}
\end{equation} For constant $\theta^A$, {\bf V} reduces to a tensor of the adjoint (or regular) representation of the
group:
\begin{equation} {\bf V}'_{\mu} = U{\bf V}_{\mu}U^{-1} \simeq {\bf V}_{\mu} + ig[\theta, {\bf V}_{\mu}]~,
\label{13}
\end{equation} which implies that
\begin{equation} V'^C_{\mu} = V^C_{\mu} - gC_{ABC}\theta^AV^B_{\mu}~,
\label{14}
\end{equation} where repeated indices are summed up.

As a consequence of Eqs. (\ref{11}) and (\ref{12}), $D_{\mu}\phi$  has the same transformation pro\-perties as $\phi$:
\begin{equation} (D_{\mu}\phi)' = U(D_{\mu}\phi)~.
\label{15}
\end{equation}

Thus ${\cal L}[\phi,D_{\mu}\phi]$ is indeed invariant under gauge transformations. In order to construct a
gauge-invariant kinetic energy term for the gauge fields $V^A$, we consider
\begin{equation} [D_{\mu},D_{\nu}] \phi =  ig\{\partial_{\mu}{\bf V}_{\nu} - \partial_{\nu}{\bf V}_{\mu} + ig[{\bf
V}_{\mu},{\bf V}_{\nu}]\}\phi \equiv ig {\bf F}_{\mu\nu}\phi~,
\label{16}
\end{equation} which is equivalent to
\begin{equation} F^A_{\mu\nu} = \partial_{\mu}V^A_{\nu} - \partial_{\nu}V^A_{\mu} - gC_{ABC}V^B_{\mu}V^C_{\nu}~.
\label{17}
\end{equation} From Eqs. (\ref{10}), (\ref{15}) and (\ref{16}) it follows that the transformation properties of
$F^A_{\mu\nu}$ are those of a tensor of the adjoint representation
\begin{equation} {\bf F}'_{\mu\nu} = U{\bf F}_{\mu\nu}U^{-1}~.
\label{18}
\end{equation} The complete Yang--Mills Lagrangian, which is invariant under gauge transformations, can be written in
the form
\begin{equation} {\cal L}_{\rm YM} = - \frac{1}{4} \sum_A F^A_{\mu\nu}F^{A\mu\nu} + {\cal L} [\phi,D_{\mu}\phi]~.
\label{19}
\end{equation}

For an Abelian theory, as for example QED, the gauge transformation reduces to
$U[\theta(x)] = {\rm exp} [ieQ\theta(x)]$, where $Q$ is the charge generator. The associated gauge field (the photon),
according to Eq. (\ref{12}), transforms as
\begin{equation} V'_{\mu} = V_{\mu} - \partial_{\mu}\theta(x)~.
\label{20}
\end{equation} In this case, the $F_{\mu\nu}$ tensor is linear in the gauge field $V_{\mu}$ so that in the absence of
matter fields the theory is free. On the other hand, in the non-Abelian case the $F^A_{\mu\nu}$ tensor contains both
linear and quadratic terms in $V^A_{\mu}$, so that the theory is non-trivial even in the absence of matter fields.

\section{The Standard Model of Electroweak Interactions}

 In this section, we summarize the structure of the standard EW Lagrangian and specify the couplings of
$W^{\pm}$ and $Z$, the intermediate vector bosons. 

For this discussion we split the Lagrangian into two parts by separating the Higgs boson couplings:
\begin{equation} {\cal L} = {\cal L}_{\rm symm} + {\cal L}_{\rm Higgs}~.
\label{21}
\end{equation}

We start by specifying ${\cal L}_{\rm symm}$, which involves only gauge bosons and fermions:
\begin{eqnarray} {\cal L}_{\rm symm} &=& -\frac{1}{4}~\sum^3_{A=1}~F^A_{\mu\nu}F^{A\mu\nu} -
\frac{1}{4}B_{\mu\nu}B^{\mu\nu} +
\bar\psi_Li\gamma^{\mu}D_{\mu}\psi_L \nonumber \\ &&+  \bar\psi_Ri\gamma^{\mu}D_{\mu}\psi_R~.
\label{22}
\end{eqnarray} This is the Yang--Mills Lagrangian for the gauge group $SU(2)\otimes U(1)$ with fermion matter fields.
Here
\begin{equation} B_{\mu\nu}  =  \partial_{\mu}B_{\nu} - \partial_{\nu}B_{\mu} \quad {\rm and} \quad F^A_{\mu\nu} =
\partial_{\mu}W^A_{\nu} - \partial_{\nu}W^A_{\mu}  - g \epsilon_{ABC}~W^B_{\mu}W^C_{\nu}
\label{23}
\end{equation} are the gauge antisymmetric tensors constructed out of the gauge field $B_{\mu}$ associated with $U(1)$,
and $W^A_{\mu}$ corresponding to the three $SU(2)$ generators; $\epsilon_{ABC}$ are the group structure constants [see
Eqs. (\ref{9})] which, for $SU(2)$, coincide with the totally antisymmetric Levi-Civita tensor (recall the familiar
angular momentum commutators). The normalization of the $SU(2)$ gauge coupling $g$ is therefore specified by
Eq.~(\ref{23}).

The fermion fields are described through their left-hand and right-hand components:
\begin{equation}
\psi_{L,R} = [(1 \mp \gamma_5)/2]\psi, \quad
\bar \psi_{L,R} = \bar \psi[(1 \pm \gamma_5)/2]~,
\label{24}
\end{equation} with $\gamma_5$ and other Dirac matrices defined as in the book by Bjorken--Drell. In particular, $\gamma^2_5
= 1, \gamma_5^{\dag} = \gamma_5$. Note that, as given in Eq. (\ref{24}),
$$
\bar\psi_L = 
\psi^{\dag}_L\gamma_0 = \psi^{\dag}[(1-\gamma_5)/2]\gamma_0 =
\bar\psi[\gamma_0(1-\gamma_5)/2]\gamma_0 = \bar \psi[(1 + \gamma_5)/2]~.
$$ The matrices $P_{\pm} = (1 \pm \gamma_5)/2$ are projectors. They satisfy the relations $P_{\pm}P_{\pm} = P_{\pm},
P_{\pm}P_{\mp} = 0, P_+ + P_- = 1$.

The sixteen linearly independent Dirac matrices can be divided into
$\gamma_5$-even and $\gamma_5$-odd according to whether they commute or anticommute with $\gamma_5$. For the
$\gamma_5$-even, we have
\begin{equation}
\bar \psi\Gamma_E \psi = \bar \psi_L\Gamma_E\psi_R + \bar \psi_R\Gamma_E\psi_L
\quad\quad (\Gamma_E \equiv 1, i\gamma_5, \sigma_{\mu\nu})~,
\label{25}
\end{equation} whilst for the $\gamma_5$-odd,
\begin{equation}
\bar \psi\Gamma_O \psi = \bar \psi_L\Gamma_O\psi_L + \bar \psi_R\Gamma_O\psi_R
\quad\quad (\Gamma_O \equiv \gamma_{\mu}, \gamma_{\mu}\gamma_5)~.
\label{26}
\end{equation} In the Standard Model (SM) the left and right fermions have different transformation properties under
the gauge group. Thus, mass terms for fermions (of the form
$\bar\psi_L\psi_R$ + h.c.) are forbidden in the symmetric limit. In particular, all $\psi_R$ are singlets in the
Minimal Standard Model (MSM). But for the moment, by
$\psi_R$ we mean a column vector, including all fermions in the theory that span a generic reducible representation of
$SU(2) \otimes U(1)$. The standard EW theory is a chiral theory, in the sense that $\psi_L$ and $\psi_R$ behave
differently under the gauge group. In the absence of mass terms, there are only vector and axial vector interactions in
the Lagrangian that have the property of not mixing $\psi_L$ and $\psi_R$. Fermion masses will be introduced, together
with
$W^{\pm}$ and $Z$ masses, by the mechanism of symmetry breaking. The covariant derivatives $D_{\mu}\psi_{L,R}$ are
explicitly given by
\begin{equation} D_{\mu}\psi_{L,R} = 
\left[ \partial_{\mu} + ig \sum^3_{A=1}~t^A_{L,R}W^A_{\mu} + ig'\frac{1}{2}Y_{L,R}B_{\mu} \right] \psi_{L,R}~,
\label{27}
\end{equation}  where $t^A_{L,R}$ and $1/2Y_{L,R}$ are the $SU(2)$ and $U(1)$ generators, respectively, in the
reducible representations $\psi_{L,R}$. The commutation relations of the $SU(2)$ generators are given by
\begin{equation} [t^A_L,t^B_L] = i~\epsilon_{ABC}t^C_L \quad {\rm and} \quad [t^A_R,t^B_R] = i \epsilon_{ABC}t^C_R~.
\label{28}
\end{equation} We use the normalization (\ref{8}) [in the fundamental representation of
$SU(2)$]. The electric charge generator $Q$ (in units of $e$, the positron charge) is given by
\begin{equation} Q = t^3_L + 1/2~Y_L = t^3_R + 1/2~Y_R~.
\label{29}
\end{equation} Note that the normalization of the $U(1)$ gauge coupling $g'$ in (\ref{27}) is now specified as a
consequence of (\ref{29}).

All fermion couplings to the gauge bosons can be derived directly from Eqs. (\ref{22}) and (\ref{27}). The
charged-current (CC) couplings are the simplest. From
\begin{eqnarray} g(t^1W^1_{\mu} + t^2W^2_{\mu}) &=& g \left\{ [(t^1 + it^2)/ \sqrt 2] (W^1_{\mu} - iW^2_{\mu})/\sqrt 2]
+ {\rm h.c.} \right\}\nonumber \\
 &= &g \left\{[(t^+W^-_{\mu})/\sqrt 2] + {\rm h.c.} \right\}~,
\label{30}
\end{eqnarray} where $t^{\pm}  = t^1 \pm it^2$ and $W^{\pm} = (W^1 \pm iW^2)/\sqrt 2$, we obtain the vertex
\begin{equation} V_{\bar \psi \psi W}  =  g \bar \psi \gamma_{\mu}\left[ (t^+_L/ \sqrt 2)(1 - \gamma_5)/2 + (t^+_R/
\sqrt 2)(1 + \gamma_5)/2 \right]
 \psi W^-_{\mu} + {\rm h.c.}
\label{31}
\end{equation}

In the neutral-current (NC) sector, the photon $A_{\mu}$ and the mediator
$Z_{\mu}$ of the weak NC are orthogonal and normalized linear combinations of
$B_{\mu}$ and $W^3_{\mu}$:
\begin{eqnarray} A_{\mu} &=& \cos \theta_WB_{\mu} + \sin \theta_WW^3_{\mu}~, \nonumber \\  Z_{\mu} &=& -\sin
\theta_WB_{\mu} + \cos \theta_W~W^3_{\mu}~.
\label{32}
\end{eqnarray} Equations (\ref{32}) define the weak mixing angle $\theta_W$. The photon is characterized by equal
couplings to left and right fermions with a strength equal to the electric charge. Recalling Eq. (\ref{29}) for the
charge matrix $Q$, we immediately obtain
\begin{equation} g~\sin \theta_W = g'\cos \theta_W = e~,
\label{33}
\end{equation} or equivalently,
\begin{equation} {\rm tg}~\theta_W = g'/g
\label{34}
\end{equation} Once $\theta_W$ has been fixed by the photon couplings, it is a simple matter of algebra to derive the
$Z$ couplings, with the result
\begin{equation}
\Gamma_{\bar \psi \psi Z} = g/(2~\cos \theta_W) \bar \psi \gamma_{\mu}
  [t^3_L(1-\gamma_5) + t^3_R(1+\gamma_5) - 2Q \sin^2\theta_W] \psi Z^{\mu}~,
\label{35}
\end{equation}  where $\Gamma_{\bar \psi \psi Z}$ is a notation for the vertex. In the MSM, $t^3_R = 0$ and $t^3_L =
\pm 1/2$. 

In order to derive the effective four-fermion interactions that are equivalent, at low energies, to the CC and NC
couplings given in Eqs. (\ref{31}) and (\ref{35}), we anticipate that large masses, as experimentally observed, are
provided for $W^{\pm}$  and $Z$ by ${\cal L}_{\rm Higgs}$.  For left--left CC couplings, when the momentum transfer
squared can be neglected with respect to
$m^2_W$ in the propagator of Born diagrams with single $W$ exchange, from Eq.~(\ref{31}) we can write
 \begin{equation} {\cal L}^{\rm CC}_{\rm eff} \simeq (g^2/8m^2_W) [ \bar \psi \gamma_{\mu}(1 - \gamma_5)t^+_L\psi][
\bar \psi
\gamma^{\mu}(1 - \gamma_5) t^-_L\psi]~.
\label{36}
\end{equation}  By specializing further in the case of doublet fields such as $\nu_e-e^-$ or $
\nu_{\mu} - \mu^-$, we obtain the tree-level relation of $g$ with the Fermi coupling constant $G_F$ measured from $\mu$
decay [see Eq. (\ref{2})]:
\begin{equation}
 \frac{G_F}{\sqrt 2} = \frac{g^2}{8m^2_W}~.
\label{37}
\end{equation} By recalling that $g~\sin \theta_W = e$, we can also cast this relation in the form
\begin{equation} m_W = \frac{\mu_{\rm Born}}{ \sin \theta_W}~,
\label{38}
\end{equation} with
\begin{equation}
\mu_{\rm Born} = \left(\frac{\pi \alpha}{\sqrt 2 G_F}\right)^{1/2} \simeq 37.2802~{\rm GeV}~,
\label{39}
\end{equation} where $\alpha$ is the fine-structure constant of QED $(\alpha \equiv e^2/4\pi = 1/137.036)$. 

In the same way, for neutral currents we obtain in Born approximation from Eq.~(\ref{35}) the effective four-fermion
interaction given by
\begin{equation} {\cal L}^{\rm NC}_{\rm eff} \simeq \sqrt 2~G_F \rho_0\bar \psi \gamma_{\mu}[...]
\psi \bar \psi \gamma^{\mu}[...] \psi~,
\label{40}
\end{equation} where
\begin{equation} [...] \equiv t^3_L(1 - \gamma_5) + t^3_R (1 + \gamma_5) - 2Q \sin^2\theta_W
\label{41}
\end{equation} and
\begin{equation}
\rho_0 = m^2_W/m^2_Z~\cos^2 \theta_W~.
\label{42}
\end{equation}

All couplings given in this section are obtained at tree level and are modified in higher orders of perturbation
theory. In particular, the relations between
$m_W$ and $\sin \theta_W$  [Eqs. (\ref{38}) and (\ref{39})] and the observed values of $\rho~(\rho = \rho_0$ at tree
level) in different NC processes, are altered by computable EW radiative corrections, as discussed in Section
6. 

The gauge-boson self-interactions can be derived from the
$F_{\mu\nu}$ term in ${\cal L}_{\rm symm}$, by using Eq. (\ref{32}) and
$W^{\pm} = (W^1 \pm iW^2)/\sqrt 2$. Defining the three-gauge-boson vertex as in Fig. 1, we obtain $(V \equiv \gamma,Z)$
\begin{equation}
\Gamma_{W^-W^+V} = ig_{W^-W^+V}[g_{\mu\nu}(q-p)_{\lambda} + g_{\mu\lambda}(p-r)_{\nu} + g_{\nu\lambda}(r-q)_{\mu}]~,
\label{43}
\end{equation} with
\begin{equation} g_{W^-W^+\gamma} = g~\sin \theta_W = e \quad {\rm and} \quad g_{W^-W^+Z} = g~\cos \theta_W~.
\label{44}
\end{equation} This form of the triple gauge vertex is very special: in general, there could be departures from the above SM
expression, even restricting us to $SU(2)\otimes U(1)$ gauge symmetric and C and P invariant couplings. In fact some small
corrections are already induced by the radiative corrections. But, in principle, more important could be the modifications
induced by some new physics effect. The experimental testing of the triple gauge vertices is presently underway at LEP2 and
limits on departures from the SM couplings have also been obtained at the Tevatron and elsewhere
(see Section~12).

\begin{figure}
\hglue 3.5cm
\epsfig{figure=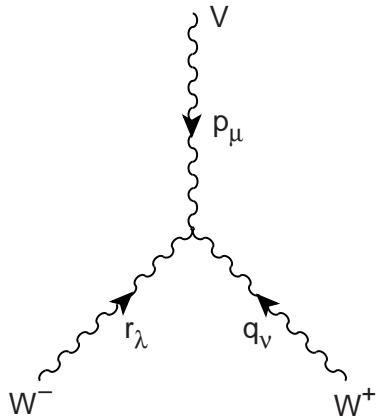, width=5cm}
\caption[]{The three-gauge boson vertex: $V=\gamma,Z$}
\end{figure}

%\begin{figure}
%\vspace{5.5 cm}
%\caption[ ]{The three-gauge boson vertex: $V=\gamma,Z$}
%\end{figure} 

We now turn to the Higgs sector of the EW Lagrangian. Here we simply review the formalism of the
Higgs mechanism applied to the EW theory. In the next section we shall make a more general and detailed
discussion of the physics of the EW symmetry breaking. The Higgs Lagrangian is specified by the gauge
principle and the requirement of renormalizability to be
\begin{equation} {\cal L}_{\rm Higgs} = (D_{\mu}\phi)^{\dag}(D^{\mu}\phi) - V(\phi^{\dag}\phi) -
\bar \psi_L \Gamma \psi_R \phi - \bar \psi_R \Gamma^{\dag} \psi_L \phi^{\dag}~,
\label{45}
\end{equation} where $\phi$ is a column vector including all Higgs fields; it transforms as a reducible representation
of the gauge group. The quantities $\Gamma$ (which include all coupling constants) are matrices that make the Yukawa
couplings invariant under the Lorentz and gauge groups. The potential $V(\phi^{\dag}\phi)$, symmetric under $SU(2)
\otimes  U(1)$, contains, at most, quartic terms in $\phi$ so that the theory is renormalizable:
\beq
V(\phi^{\dag}\phi)=-\frac{1}{2}\mu^2\phi^{\dag}\phi+\frac{1}{4}\lambda(\phi^{\dag}\phi)^2\label{44a}
\eeq

As discussed in the next section, spontaneous symmetry
breaking is induced if the minimum of V " which is the classical analogue of the quantum mechanical vacuum state (both
are the states of minimum energy) " is obtained for non-vanishing $\phi$ values. Precisely, we denote the vacuum
expectation value (VEV) of $\phi$, i.e. the position of the minimum, by $v$:
\begin{equation}
\langle 0 |\phi (x)|0 \rangle = v \not= 0~.
\label{46}
\end{equation}

The fermion mass matrix is obtained from the Yukawa couplings by replacing $\phi (x)$ by $v$:
\begin{equation} M = \bar \psi_L~{\cal M} \psi_R + \bar \psi_R {\cal M}^{\dag}\psi_L~,
\label{47}
\end{equation} with
\begin{equation} {\cal M} = \Gamma \cdot v~.
\label{48}
\end{equation} In the MSM, where all left fermions $\psi_L$ are doublets and all right fermions $\psi_R$ are singlets,
only Higgs doublets can contribute to fermion masses. There are enough free couplings in $\Gamma$, so that one single
complex Higgs doublet is indeed sufficient to generate the most general fermion mass matrix. It is important to observe
that by a suitable change of basis we can always make the matrix ${\cal M}$ Hermitian, $\gamma_5$-free, and diagonal. In
fact, we can make separate unitary transformations on $\psi_L$ and $\psi_R$ according to
\begin{equation}
\psi'_L = U\psi_L, \quad \psi'_R = V\psi_R
\label{49}
\end{equation} and consequently
\begin{equation} {\cal M} \rightarrow {\cal M}' = U^{\dag}{\cal M}V~.
\label{50}
\end{equation} This transformation does not alter the general structure of the fermion couplings in ${\cal L}_{\rm
symm}$.

 If only one Higgs doublet is present, the change of basis that makes ${\cal M}$ diagonal will at the same time
diagonalize also the fermion--Higgs Yukawa couplings. Thus, in this case, no flavour-changing neutral Higgs exchanges
are present. This is not true, in general, when there are several Higgs doublets. But one Higgs doublet for each
electric charge sector i.e. one doublet coupled only to $u$-type quarks, one doublet to $d$-type quarks, one doublet to
charged leptons would also be all right, because the mass matrices of fermions with different charges are
diagonalized separately. For several Higgs doublets in a given charge sector it is also possible to generate CP
violation by complex phases in the Higgs couplings. In the presence of six quark flavours, this CP-violation mechanism is
not necessary. In fact, at the moment, the simplest model with only one Higgs doublet seems adequate for describing all
observed phenomena.

We now consider the gauge-boson masses and their couplings to the Higgs. These effects are induced by the
$(D_{\mu}\phi)^{\dag}(D^{\mu}\phi)$ term in
${\cal L}_{\rm Higgs}$ [Eq. (\ref{45})], where
\begin{equation} D_{\mu}\phi = \left[ \partial_{\mu} + ig \sum^3_{A=1} t^AW^A_{\mu} + ig'(Y/2)B_{\mu} \right] \phi~.
\label{51}
\end{equation} Here $t^A$ and $1/2Y$ are the $SU(2) \otimes U(1)$ generators in the reducible representation spanned by
$\phi$. Not only doublets but all non-singlet Higgs representations can contribute to gauge-boson masses. The condition
that the photon remains massless is equivalent to the condition that the vacuum is electrically neutral:
\begin{equation} Q|v\rangle = (t^3 + \frac{1}{2}Y)|v \rangle = 0~.
\label{52}
\end{equation} The charged $W$ mass is given by the quadratic terms in the $W$ field arising from
${\cal L}_{\rm Higgs}$, when $\phi (x)$ is replaced by $v$. We obtain
\begin{equation} m^2_WW^+_{\mu}W^{- \mu} = g^2|(t^+v/ \sqrt 2)|^2 W^+_{\mu}W^{- \mu}~,
\label{53}
\end{equation} whilst for the $Z$ mass we get [recalling Eq. (\ref{32})]
\begin{equation}
\frac{1}{2}m^2_ZZ_{\mu}Z^{\mu} = |[g \cos \theta_Wt^3 - g' \sin
\theta_W(Y/2)]v|^2Z_{\mu}Z^{\mu}~,
\label{54}
\end{equation} where the factor of 1/2 on the left-hand side is the correct normalization for the definition of the
mass of a neutral field. By using Eq. (\ref{52}), relating the action of $t^3$ and $1/2Y$ on the vacuum $v$, and Eqs.
(\ref{34}), we obtain
\begin{equation}
\frac{1}{2}m^2_Z = (g \cos \theta_W + g' \sin \theta_W)^2 |t^3v|^2 =
\left(\frac{g^2}{\cos^2 \theta_W}\right)|t^3v|^2~.
\label{55}
\end{equation} For Higgs doublets
\begin{equation}
\phi = \pmatrix { \phi^+ \cr
\phi^0}, \quad v = \pmatrix{ 0 \cr v}~, 
\label{56}
\end{equation} we have
\begin{equation} |t^+v|^2 = v^2, \quad |t^3v|^2 = 1/4v^2~,
\label{57}
\end{equation} so that
\begin{equation}
 m^2_W = \frac{1}{2}g^2v^2, \quad m^2_Z =
 \frac{1}{2}~\frac{g^2v^2}{\cos^2\theta_W}~.
\label{58}
\end{equation} Note that by using Eq. (\ref{37}) we obtain
\begin{equation} v = 2^{-3/4}G^{-1/2}_F = 174.1~{\rm GeV}~.
\label{59}
\end{equation} It is also evident that for Higgs doublets
\begin{equation}
\rho_0 = m^2_W/m^2_Z \cos^2\theta_W = 1~.
\label{60}
\end{equation}

This relation is typical of one or more Higgs doublets and would be spoiled by the existence of Higgs triplets etc. In
general,
\begin{equation}
\rho_0 = \frac{\sum_i((t_i)^2 - (t^3_i)^2 + t_i ) v^2_i}{ \sum
_i2(t^3_i)^2v^2_i}~,
\label{61}
\end{equation} 
for several Higgses with VEVs $v_i$, weak isospin $t_i$, and $z$-component $t^3_i$. These results are
valid at the tree level and are modified by calculable EW radiative corrections, as discussed in Section 6.

In the minimal version of the SM only one Higgs doublet is present. Then the fermion--Higgs couplings are in proportion to
the fermion masses. In fact, from the Yukawa couplings $g_{\phi
\bar f f}(\bar f_L \phi f_R + h.c.)$, the mass $m_f$ is obtained by replacing
$\phi$ by $v$, so that $ m_f = g_{\phi \bar f f} v $. In the minimal SM
three out of the four Hermitian fields are removed from the physical spectrum by
the Higgs mechanism and become the longitudinal modes of $W^+, W^-$, and $Z$. The fourth neutral Higgs is physical and
should be found. If more doublets are present, two more charged and two more neutral Higgs scalars should be around for
each additional doublet.

The couplings of the physical Higgs $H$ to the gauge bosons can be simply obtained from ${\cal L}_{\rm
Higgs}$, by the replacement
\begin{equation}
\phi(x) = \pmatrix{ \phi^+(x) \cr
\phi^0(x)} \rightarrow 
\pmatrix {0 \cr v + (H/\sqrt2)}~,
\label{62}
\end{equation}
 [so that $(D_{\mu}\phi)^{\dag}(D^{\mu}\phi) = 1/2(\partial_{\mu}H)^2 + ...]$, with the result
\begin{eqnarray}
 {\cal L} [H,W,Z] &=& g^2 \left(\frac{v}{\sqrt 2}\right)
 W^+_{\mu}W^{-\mu} H + \left(\frac{g^2 }{4}\right)W^+_{\mu}W^{-\mu}H^2 \nonumber \\ &&
+ \left[ \frac{g^2vZ_{\mu}Z^{\mu}}{2 \sqrt 2
\cos^2\theta_W}\right]H \nonumber \\ 
&&+ \left[\frac{g^2}{8 \cos^2\theta_W}\right]Z_{\mu}Z^{\mu}H^2~.
\label{63}
\end{eqnarray}

In the minimal SM the Higgs mass $m^2_H\sim \lambda v^2$ is of order of the weak scale v. We will discuss in sect. 8 the
direct experimental limit on $m_H$ from LEP, which is $m_H\gappeq m_Z$. We shall also see in sect.12 , that, if there is no
physics beyond the SM up to a large scale $\Lambda$, then, on theoretical grounds, $m_H$ can only be within a narrow range
between 135 and 180 GeV. But the interval is enlarged if there is new physics nearby. Also the lower limit depends
critically on the assumption of only one doublet. The dominant decay mode of the Higgs is in the $b \bar b$ channel below
the WW threshold, while the $W^+W^-$ channel is dominant for sufficiently large $m_H$. The width is small below the WW
threshold, not exceeding a few MeV, but increases steeply beyond the threshold, reaching the asymptotic value of $\Gamma\sim
1/2 m^3_H$ at large $m_H$, where all energies are in TeV.

\section{The Higgs Mechanism}

The gauge symmetry of the Standard Model was difficult to discover because it is well hidden in nature. The only
observed gauge boson that is massless is the photon. The gluons are presumed massless but are unobservable because of
confinement, and the $W$ and $Z$ weak bosons carry a heavy mass. Actually a major difficulty in unifying weak and
electromagnetic interactions was the fact that e.m. interactions have infinite range $(m_{\gamma} = 0)$, whilst the weak
forces have a very short range, owing to
$m_{W,Z} \not= 0$.

The solution of this problem is in the concept of spontaneous symmetry breaking, which was borrowed from statistical
mechanics. 

Consider a ferromagnet at zero magnetic field in the Landau--Ginzburg approximation. The free energy in terms of the
temperature $T$ and the magnetization {\bf M} can be written as
\begin{equation} F({\bf M}, T) \simeq F_0(T) + 1/2~\mu^2(T){\bf M}^2 + 1/4~\lambda(T)({\bf M}^2)^2 + ...~.
\label{64}
\end{equation} This is an expansion which is valid at small magnetization.  The neglect of terms of higher order in
$\vec M^2$ is the analogue in this context of the renormalizability criterion. Also, $\lambda(T) > 0$ is assumed for
stability; $F$ is invariant under rotations, i.e. all directions of {\bf M} in space are equivalent. The minimum
condition for $F$ reads
\begin{equation}
\partial F/\partial M = 0, \quad [\mu^2(T) + \lambda(T){\bf M}^2]{\bf M} = 0~.
\label{65}
\end{equation} There are two cases. If $\mu^2 > 0$, then the only solution is ${\bf M} = 0$, there is no magnetization,
and the rotation symmetry is respected. If $\mu^2 < 0$, then another solution appears, which is
\begin{equation} |{\bf M}_0|^2 = -\mu^2/\lambda~.
\label{66}
\end{equation} The direction chosen by the vector ${\bf M}_0$ is a breaking of the rotation symmetry. The critical
temperature $T_{\rm crit}$ is where $\mu^2(T)$ changes sign:
\begin{equation}
\mu^2(T_{\rm crit}) = 0~.
\label{67}
\end{equation} It is simple to realize that the Goldstone theorem holds. It states that when spontaneous symmetry
breaking takes place, there is always a zero-mass mode in the spectrum. In a classical context this can be proven as
follows. Consider a Lagrangian
\begin{equation} {\cal L} = |\partial_{\mu}\phi|^2 - V(\phi)
\label{68}
\end{equation} symmetric under the infinitesimal transformations
\begin{equation}
\phi \rightarrow  \phi' = \phi + \delta \phi, \quad
\delta \phi_i = i \delta \theta t_{ij}\phi_j~.
\label{69}
\end{equation} The minimum condition on $V$ that identifies the equilibrium position (or the ground state in quantum
language) is
\begin{equation} (\partial V/\partial \phi_i)(\phi_i = \phi^0_i) = 0~.
\label{70}
\end{equation} The symmetry of $V$ implies that
\begin{equation}
\delta V = (\partial V/\partial \phi_i)\delta \phi_i = i \delta \theta(\partial V/\partial \phi_i)t_{ij}\phi_j = 0~.
\label{71}
\end{equation} By taking a second derivative at the minimum $\phi_i = \phi^0_i$ of the previous equation, we obtain
\begin{equation}
\partial^2V/\partial \phi_k\partial \phi_i (\phi_i =
\phi^0_i)t_{ij}\phi^0_i + \frac{\partial V}{\partial \phi_i} (\phi_i =
\phi^0_i)t_{ik} = 0~.
\label{72}
\end{equation} The second term vanishes owing to the minimum condition, Eq. (\ref{70}). We then find
\begin{equation}
\partial^2V/\partial \phi_k\partial \phi_i ~(\phi_i = \phi^0_i)t_{ij}\phi^0_j = 0~.
\label{73}
\end{equation} The second derivatives $M^2_{ki} = (\partial^2V/\partial \phi_k \partial
\phi_i)(\phi_i = \phi^0_i)$ define the squared mass matrix. Thus the above equation in matrix notation can be read as
\begin{equation} M^2 t\phi^0 = 0~,
\label{74}
\end{equation} which shows that if the vector $(t\phi^0)$ is non-vanishing, i.e. there is some generator that shifts
the ground state into some other state with the same energy, then $t \phi^0$ is an eigenstate of the squared mass
matrix with zero eigenvalue. Therefore, a massless mode is associated with each broken generator.

When spontaneous symmetry breaking takes place in a gauge theory, the massless Goldstone mode exists, but it is
unphysical and disappears from the spectrum. It becomes, in fact, the third helicity state of a gauge boson that takes
mass. This is the Higgs mechanism. Consider, for example, the simplest Higgs model described by the Lagrangian
\begin{equation} {\cal L} = -\frac{1}{4}~F^2_{\mu\nu} + |(\partial_{\mu} - ieA_{\mu})\phi|^2 +
\frac{1}{2} \mu^2 \phi^*\phi - (\lambda/4)(\phi^*\phi)^2~.
\label{75}
\end{equation} Note the `wrong' sign in front of the mass term for the scalar field $\phi$, which is necessary for the
spontaneous symmetry breaking to take place. The above Lagrangian is invariant under the $U(1)$ gauge symmetry
\begin{equation} A_{\mu} \rightarrow A'_{\mu} = A_{\mu} - (1/e)\partial_{\mu}\theta(x), \quad
\phi \rightarrow \phi' = \phi ~{\rm exp}[i\theta(x)]~.
\label{76}
\end{equation} Let $\phi^0 = v \not= 0$, with $v$ real, be the ground state that minimizes the potential and induces
the spontaneous symmetry breaking. Making use of gauge invariance, we can make the change of variables
\begin{eqnarray} &&\phi(x) \rightarrow (1/\sqrt 2)[\rho(x) + v]~{\rm exp}[i \zeta(x)/v]~, \nonumber \\ &&A_{\mu}(x)
\rightarrow A_{\mu} - (1/ev)\partial_{\mu}  \zeta(x).
\label{77}
\end{eqnarray} Then $\rho = 0$ is the position of the minimum, and the Lagrangian becomes
\begin{equation} {\cal L} = -\frac{1}{4}F^2_{\mu\nu} + \frac{1}{2}e^2v^2A^2_{\mu} + \frac{1}{2} e^2\rho^2A^2_{\mu} +
e^2\rho vA^2_{\mu} + {\cal L}(\rho)~.
\label{78}
\end{equation} The field $\zeta(x)$, which corresponds to the would-be Goldstone boson, disappears, whilst the mass
term $\frac{1}{2}e^2v^2A^2_{\mu}$ for $A_{\mu}$ is now present; $\rho$ is the massive Higgs particle.

The Higgs mechanism is realized in well-known physical situations. For a superconductor in the Landau--Ginzburg
approximation the free energy can be written as
\begin{equation} F = F_0 + \frac{1}{2}{\bf B}^2 + |({\bf \nabla} - 2ie{\bf A})\phi|^2/4m -
\alpha|\phi|^2 + \beta|\phi|^4~.
\label{79}
\end{equation} 

Here {\bf B} is the magnetic field, $|\phi|^2$ is the Cooper pair $(e^-e^-)$ density, 2$e$ and 2$m$ are the charge and
mass of the Cooper pair. The 'wrong' sign of $\alpha$ leads to $\phi \not= 0$ at the minimum. This is precisely the
non-relativistic analogue of the Higgs model of the previous example. The Higgs mechanism implies the absence of
propagation of massless phonons (states with dispersion relation ~$\omega = kv$ with constant $v$). Also the mass term
for {\bf A} is manifested by the exponential decrease of {\bf B} inside the superconductor (Meissner effect).

\section{The CKM Matrix}

Weak charged currents are the only tree level interactions in the SM that change flavour: by emission of a W an
up-type quark is turned into a  down-type quark, or a $\nu_l$ neutrino is turned into a $l^-$
charged lepton (all fermions are letf-handed). If we start from an up quark that is a mass
eigenstate, emission of a W turns it into a down-type quark state d' (the weak isospin partner of
u) that in general is not a mass eigenstate. In general, the mass eigenstates and the weak
eigenstates do not coincide and a unitary transformation connects the two sets:
\beq
\left(\matrix{d^\prime\cr s^\prime\cr b^\prime}\right)=V\left(\matrix{d\cr s\cr b}\right)\label{km1}
\eeq
V is the Cabibbo-Kobayashi-Maskawa matrix.
Thus in terms of mass eigenstates the charged weak current of quarks is of the form:
\beq
J^+_{\mu}\propto\bar u \gamma_{\mu}(1-\gamma_5)t^+ Vd 
\label{km2}
\eeq
Since V is unitary (i.e. $VV^\dagger=V^\dagger V=1$) and commutes with $T^2$, $T_3$ and Q (because all d-type quarks
have the same isospin and charge) the neutral current couplings are diagonal both in the primed and unprimed basis (if
the Z down-type quark current is abbreviated as $\bar d^\prime \Gamma d^\prime$ then by changing basis we get $\bar d
V^\dagger \Gamma V d$ and V and $\Gamma$ commute because, as seen from eq.(\ref{41}), $\Gamma$ is made of Dirac
matrices and $T_3$ and Q generator matrices). This is the GIM mechanism that ensures natural flavour conservation of
the neutral current couplings at the tree level. 

For N generations of quarks, V is a NxN unitary matrix that depends on $N^2$ real numbers ($N^2$ complex entries with
$N^2$ unitarity constraints). However, the $2N$ phases of up- and down-type quarks are not observable. Note that an
overall phase drops away from the expression of the current in eq.(\ref{km2}), so that only $2N-1$ phases can affect V.
In total, V depends on $N^2-2N+1=(N-1)^2$ real physical parameters. A similar counting gives $N(N-1)/2$ as the number of
independent parameters in an orthogonal NxN matrix. This implies that in V we have $N(N-1)/2$ mixing angles and
$(N-1)^2-N(N-1)/2$ phases: for $N=2$ one mixing angle (the Cabibbo angle) and no phase, for $N=3$ three angles and one
phase etc. 

Given the experimental near diagonal structure of V a convenient parametrisation is the one proposed by
Maiani. One starts from the definition:
\beq
\vert d'\rangle=c_{13}\vert d_C\rangle+s_{13} e^{-i\phi}\vert b\rangle
\label{km3}
\eeq
where $c_{13}\equiv cos\theta_{13}$, $s_{13}\equiv sin\theta_{13}$ (analogous shorthand notations will be used in the
following), $d_C$ is the Cabibbo down quark and  $\theta_{12}\equiv\theta_C$ is the
Cabibbo angle (experimentally $s_{12}\equiv\lambda\sim 0.22$).
\beq
\vert d_C\rangle=c_{12}\vert d\rangle+s_{12} \vert s\rangle\label{km4}
\eeq
Note that in a four quark model the Cabibbo angle fixes both the ratio of couplings $(u\rightarrow
d)/(\nu_e\rightarrow e)$ and the ratio of $(u\rightarrow
d)/(u\rightarrow s)$. In a six quark model one has to choose which to keep as a definition of the Cabibbo angle.
Here the second definition is taken and, in fact the $u\rightarrow d$ coupling is given by $V_{ud}=c_{13} c_{12}$ so
that it is no longer specified by $\theta_{12}$ only. Also note that we can certainly fix the phases of u, d, s so
that a real coefficient appears in front of $d_C$ in eq.(\ref{km3}). We now choose a basis of two orthonormal vectors,
both orthogonal to $\vert d'\rangle$:
\beq
\vert s_C\rangle=-s_{12}\vert d\rangle+c_{12} \vert s\rangle,~~~~~~~\vert v\rangle=-s_{13} e^{i\phi}\vert
d_C\rangle+c_{13} \vert b\rangle\label{km5}
\eeq 
Here $\vert s_C\rangle$ is the Cabibbo s quark. Clearly s' and b' must be othonormal superpositions of the above base
vectors defined in terms of an angle $\theta_{23}$:
\beq
\vert s'\rangle=c_{23}\vert s_C\rangle+s_{23} \vert v\rangle,~~~~~~\vert b'\rangle=-s_{23}\vert
s_C\rangle+c_{23} \vert v\rangle\label{km6}
\eeq 
The general expression of $V_{ij}$ can be obtained from the above equations. But a considerable notational
simplification is gained if one takes into account that from experiment we know that $s_{12}\equiv\lambda$, $s_{23}\sim
o(\lambda^2)$ and 
$s_{13}\sim o(\lambda^3)$ are increasingly small and of the indicated orders of magnitude. Thus, following Wolfenstein
one can set:
\beq
s_{12}\equiv\lambda,~~~~~~~~s_{23}=A\lambda^2,~~~~~~~~s_{13}e^{-i\phi}=A\lambda^3(\rho-i\eta)\label{km7}
\eeq
As a result, by neglecting terms of higher order in $\lambda$ one can write down:
\beq
V= 
\left[\matrix{
V_{ud}&V_{us}&V_{ub} \cr
V_{cd}&V_{cs}&V_{cb}\cr
V_{td}&V_{ts}&V_{tb}     } 
\right ]~~~\sim~~~\left[\matrix{
1-\frac{\lambda^2}{2}&\lambda&A\lambda^3(\rho-i\eta) \cr
-\lambda&1-\frac{\lambda^2}{2}&A\lambda^2\cr
A\lambda^3(1-\rho-i\eta)&-A\lambda^2&1     } 
\right ].
\label{km8}
\eeq 
Indicative values of the CKM parameters as obtained from experiment are (a survey of the current status of the CKM
parameters can be found in ref.\cite{pdg}):
\bea
\lambda=0.2196\pm0.0023\nonumber\\
A=0.82\pm0.04\nonumber\\
\sqrt{\rho^2+\eta^2}=0.4\pm0.1;~~~~~ \eta\sim 0.3\pm0.2\label{km9}
\eea

In the SM the non vanishing of the $\eta$ parameter is the only source of CP violation. The most direct and solid evidence for
$\eta$ non vanishing is obtained from the measurement of $\epsilon$ in K decay.
Unitarity of the CKM matrix V implies relations of the form $\sum_a V_{ba}V^*_{ca}=\delta_{bc}$. In most cases these
relations do not imply particularly instructive constraints on the Wolfenstein parameters. But when the three terms in the sum
are of comparable magnitude we get interesting information. The three numbers which must add to zero form a closed triangle in
the complex plane, with sides of comparable length. This is the case for the t-u triangle (Bjorken triangle) shown in
fig.2: 
\beq
V_{td}V^*_{ud}+V_{ts}V^*_{us}+V_{tb}V^*_{ub}=0\label{km10}
\eeq
%%%%%%%%%%%%%%%%%%%%%%%%%%%%%%%%%%%%%%%%%
\begin{figure}
\hglue 3.5cm
\epsfig{figure=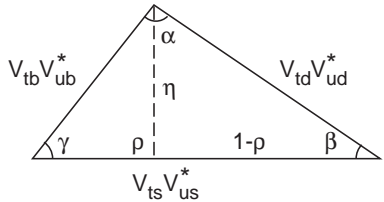, width=5cm}
\caption[]{The Bjorken triangle corresponding to eq.(\ref{km10})}
\end{figure}
%%%%%%%%%%%%%%%%%%%%%%%%%%%%%%%%%%%%%%%%%%%%%%
All terms are of order $\lambda^3$. For $\eta$=0 the triangle would flatten down to vanishing area. In fact
the area of the triangle, J of order $J\sim \eta A^2 \lambda^6$, is the Jarlskog invariant (its value is independent of the
parametrization). In the SM all CP violating observables must be proportional to J. 

We have only discussed flavour mixing for quarks. But, clearly, if neutrino masses exist, as indicated by neutrino
oscillations, then a similar mixing matrix must also be introduced in the leptonic sector
(see section~10.2). 

\section{Renormalisation and Higher Order Corrections}

The Higgs mechanism gives masses to the Z, the $W^\pm$ and to fermions while the Lagrangian density is still
symmetric. In particular the gauge Ward identities and the conservation of the gauge currents are preserved. The validity of
these relations is an essential ingredient for renormalisability. For example the massive gauge boson propagator would have
a bad ultraviolet behaviour:
\beq
W_{\mu\nu}=\frac{-g_{\mu\nu}+\frac{q_\mu q_\nu}{m^2_W}}{q^2-m^2_W}\label{prop}
\eeq
But if the propagator is sandwiched between conserved currents $J_\mu$ the bad terms in $q_\mu q_\nu$ give a vanishing
contribution because $q_\mu J^\mu=0$ and the high energy behaviour is like for a scalar particle and compatible with
renormalisation. 

The fondamental theorem that in general a gauge theory with spontaneous symmetry breaking and the Higgs
mechanism is renormalisable was proven by 't Hooft. For a chiral theory like the SM an additional complication arises from
the existence of chiral anomalies. But this problem is avoided in the SM because the quantum numbers of the quarks and
leptons in each generation imply a remarkable (and apparently miracoulous) cancellation of the anomaly, as originally
observed by Bouchiat, Iliopoulos and Meyer. In quantum field theory one encounters an anomaly when a symmetry of the
classical lagrangian is broken by the process of quantisation, regularisation and renormalisation of the theory. For
example, in massless QCD there is no mass scale in the classical lagrangian. Thus one would predict that dimensionless
quantities in processes with only one large energy scale Q cannot depend on Q and must be constants. As well known this
naive statement is false. The process of regularisation and renormalisation necessarily introduces an energy scale which is
essentially the scale where renormalised quantities are defined. For example the renormalised coupling must be defined from
the vertices at some scale. This scale
$\mu$ cannot be zero because of infrared divergences. The scale $\mu$ destroys scale invariance because dimensionless
quantities can now depend on $Q/\mu$. The famous $\Lambda_{QCD}$ parameter is a tradeoff of $\mu$ and leads to scale
invariance breaking. Of direct relevance for the EW theory is the
Adler-Bell-Jackiw chiral anomaly. The classical lagrangian
of a theory with massless fermions is invariant under a U(1) chiral transformations
$\psi\prime=e^{i\gamma_5\theta}\psi$. The associated axial Noether current is conserved at the classical level. But, at the
quantum level, chiral symmetry is broken due to the ABJ anomaly and the current is not conserved. The chiral breaking is
introduced by a clash between chiral symmetry, gauge invariance and the regularisation procedure. The anomaly is generated
by triangular fermion loops with one axial and two vector vertices (fig.3). For neutral currents (Z and
$\gamma$) the axial coupling is proportional to the 3rd component of weak isospin $t_3$, while vector couplings are
proportional to a linear combination of
$t_3$ and the electric charge Q. Thus in order for the chiral anomaly to vanish all traces of the form $tr\{t_3QQ\}$,
$tr\{t_3t_3Q\}$, $tr\{t_3t_3t_3\}$ (and also  $tr\{t_+t_-t_3\}$etc., when charged currents are included) must vanish, where
the trace is extended over all fermions in the theory that can circulate in the loop. Now all these
traces happen to vanish for each fermion family separately. For example take $tr\{t_3QQ\}$. In one family there are, with
$t_3=+1/2$, three colours of up quarks with charge $Q=+2/3$ and one neutrino with $Q=0$ and, with $t_3=-1/2$, three colours
of down quarks with charge $Q=-1/3$ and one $l^-$ with $Q=-1$. Thus we obtain $tr\{t_3QQ\}=1/2~3~4/9-1/2~3~1/9-1/2~1=0$.
This impressive cancellation suggests an interplay among weak isospin, charge and colour quantum numbers which appears as a
miracle from the point of view of the low energy theory but is more understandable from the point of view of the high energy
theory. For example in GUTs there are similar relations where charge quantisation and colour are related: in the 5 of SU(5)
we have the content $(d,d,d,e^+,\bar\nu)$ and the charge generator has a vanishing trace in each SU(5) representation (the
condition of unit determinant, represented by the letter S in the SU(5) group name, translates into zero trace for the
generators). Thus the charge of d quarks is -1/3 of the positron charge because there are three colours.
%%%%%%%%%%%%%%%%%%%%%%%%%%%%%%%%%
\begin{figure}
\hglue 4.0cm
\epsfig{figure=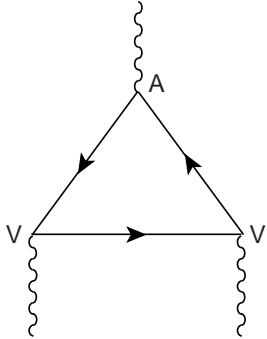, width=3.5cm}
\caption[]{Triangle diagram that generates the ABJ anomaly}
\end{figure}
%%%%%%%%%%%%%%%%%%%%%%%%%%%%%%%%%%

Since the SM theory is renormalisable higher order perturbative corrections can be reliably computed. Radiative corrections
are very important for precision EW tests. The SM inherits all successes of the old V-A theory of charged currents and of
QED. Modern tests focus on neutral current processes, the W mass and the measurement of triple gauge vertices. For Z physics
and the W mass the state of the art computation of radiative corrections include the complete one loop diagrams and selected
dominant two loop corrections. In addition some resummation techniques are also implemented, like Dyson resummation of vacuum
polarisation functions and important renormalisation group improvements for large QED and QCD logarithms. We now discuss in
more detail sets of large radiative corrections which are particularly
significant~\cite{radcorr}.

A set of important quantitative contributions to the radiative corrections arise from large logarithms [e.g. terms of the
form $(\alpha/\pi ~{\rm ln}~(m_Z/m_{f_\ell}))^n$ where $f_{\ell}$ is a light fermion]. The sequences of leading and
close-to-leading logarithms are fixed by well-known and consolidated techniques ($\beta$ functions, anomalous
dimensions, penguin-like diagrams, etc.). For example, large logarithms dominate the running of
$\alpha$ from $m_e$, the electron mass, up to $m_Z$. Similarly large logarithms of the form $[\alpha/\pi~{\rm
ln}~(m_Z/\mu)]^n$ also enter, for example, in the relation between $\sin^2\theta_W$ at the scales $m_Z$ (LEP, SLC) and $\mu$
(e.g. the scale of low-energy neutral-current experiments). Also, large logs from initial state radiation dramatically distort
the line shape of the Z resonance as observed at LEP and SLC and must be accurately taken into account in the measure of
the Z mass and total width.

For example, a considerable amount of work has deservedly been devoted to the theoretical study of the $Z$ line-shape. The
present experimental accuracy on $m_Z$ obtained at LEP is $\delta m_Z = \pm 2.1$~MeV (see table 1 , sect.7). This small
error  was obtained by a precise calibration of the LEP energy scale achieved by taking advantage of the transverse
polarization of the beams and implementing a sophisticated resonant spin depolarization method. Similarly, a
measurement of the total width to an accuracy $\delta \Gamma = \pm 2.4$~MeV has by now been achieved. The prediction of the
Z line-shape in the SM to such an accuracy has posed a formidable challenge to theory, which has been
successfully met. For the inclusive process $e^+e^- \rightarrow f \bar fX$, with $f \not= e$ (for simplicity, we leave
Bhabha scattering aside) and $X$ including $\gamma$'s and gluons, the physical cross-section can be written in the form of a
convolution
\cite{radcorr}: 
\begin{equation} \sigma(s) = \int^1_{z_0} dz~\hat \sigma(zs)G(z,s)~,
 \label{92}
\end{equation}  where $\hat \sigma$ is the reduced cross-section, and $G(z,s)$ is the radiator function that describes the
effect of initial-state radiation; $\hat \sigma$ includes the purely weak corrections, the effect of final-state radiation
(of both $\gamma$'s and gluons), and also non-factorizable terms (initial- and final-state radiation interferences, boxes,
etc.) which, being small, can be treated in lowest order and effectively absorbed in a modified $\hat \sigma$. The radiator
$G(z,s)$ has an expansion of the form
\begin{eqnarray} G(z,s) & = &
\delta(1-z) + \alpha /\pi(a_{11}L + a_{10}) + (\alpha/\pi)^2 (a_{22}L^2 + a_{11}L + a_{20}) \nonumber \\ && +~... +
(\alpha/\pi)^n~\sum^n_{i=0} a_{ni}L^i~,
\label{131}
\end{eqnarray} where $L = {\rm ln}~s/m^2_e \simeq 24.2$ for $\sqrt s \simeq m_Z$. All first- and second-order terms are known
exactly. The sequence of leading and next-to-leading logs can be exponentiated (closely following the formalism
of structure functions in QCD). For $m_Z \approx 91$~GeV, the convolution displaces the peak by  +110~MeV, and reduces it by
a factor of about 0.74. The exponentiation is important in that it amounts to a shift of about 14~MeV in the peak position.
 
A very remarkable class of contributions among the one loop EW radiative corrections are those terms that increase
quadratically with the top mass.  The sensitivity
of radiative corrections to $m_t$ arises from the existence of these terms. The quadratic dependence on
$m_t$ (and on other possible widely broken isospin multiplets from new physics) arises because, in spontaneously broken
gauge theories, heavy loops do not decouple. On the contrary, in QED or QCD, the running of
$\alpha$ and $\alpha_s$ at a scale $Q$ is not affected by heavy quarks with mass
$M \gg Q$. According to an intuitive decoupling theorem\cite{ppi}
, diagrams with heavy virtual particles of mass $M$ can be
ignored at $Q \ll M$ provided that the couplings do not grow with $M$ and that the theory with no heavy particles is still
renormalizable. In the spontaneously broken EW gauge theories both requirements are violated. First, one important difference
with respect to unbroken gauge theories is in the longitudinal modes of weak gauge bosons. These modes are generated by the
Higgs mechanism, and their couplings grow with masses (as is also the case for the physical Higgs couplings). Second the
theory without the top quark is no more renormalisable because the gauge symmetry is broken
since the doublet (t,b) would
not be complete (also the chiral anomaly would not be completely cancelled). With the observed value of
$m_t$ the quantitative importance of the terms of order $G_Fm^2_t/4\pi^2\sqrt{2}$ is substancial but not dominant (they are
enhanced by a factor $m^2_t/m^2_W\sim 5$ with respect to ordinary terms). Both the large logarithms and the
$G_Fm^2_t$ terms have a simple structure and are to a large extent universal, i.e. common to a wide class of processes. In
particular the $G_Fm^2_t$ terms appear in vacuum polarisation diagrams which are universal and in the $Z\rightarrow b \bar b$
vertex which is not (this vertex is connected with the top quark which runs in the loop, while other types of heavy particles
could  in principle also contribute to vacuum polarisation diagrams). Their study is important for an understanding of the
pattern of radiative corrections. One can also derive approximate formulae (e.g. improved Born approximations), which can be
useful in cases where a limited precision may be adequate.  More in general, another very important consequence of non
decoupling is that precision tests of the electroweak theory may be sensitive to new physics even if the new particles are
too heavy for their direct production.

While radiative corrections are quite sensitive to the top mass, they are unfortunately much less dependent on the Higgs
mass. If they were sufficiently sensitive by now we would precisely know the mass of the Higgs. But the dependence of one
loop diagrams on
$m_H$ is only logarithmic:
$\sim G_Fm^2_W
\log(m^2_H/m^2_W)$. Quadratic terms $\sim G^2_Fm^2_H$ only appear at two loops and are too small to be important. The
difference with the top case is that the difference $m^2_t-m^2_b$ is a direct breaking of the gauge symmetry that
already affects the one loop corrections, while the Higgs couplings are "custodial" SU(2) symmetric in lowest order.

The basic tree level relations:
\beq
\frac{g^2}{8m^2_W}=\frac{G_F}{\sqrt{2}},~~~~~~g^2\sin^2\theta_W=e^2=4\pi\alpha\label{bb1}
\eeq
can be combined into
\beq
\sin^2\theta_W=\frac{\pi\alpha}{\sqrt{2}G_Fm^2_W}\label{bb2}
\eeq
A different definition of $\sin^2\theta_W$ is from the gauge boson masses:
\beq
\frac{m^2_W}{m^2_Z\cos^2\theta_W}=\rho_0=1~~~\Longrightarrow~~~\sin^2\theta_W=1-\frac{m^2_W}{m^2_Z}\label{bb3}
\eeq
where $\rho_0=1$ assuming that there are only Higgs doublets. The last two relations can be put into the convenient form
\beq
(1-\frac{m^2_W}{m^2_Z})\frac{m^2_W}{m^2_Z}=\frac{\pi\alpha}{\sqrt{2}G_Fm^2_Z}\label{bb4}
\eeq
These relations are modified by radiative corrections:
\bea
(1-\frac{m^2_W}{m^2_Z})\frac{m^2_W}{m^2_Z}=\frac{\pi\alpha(m_Z)}{\sqrt{2}G_Fm^2_Z}\frac{1}{1-\Delta r_W}\nonumber\\
\frac{m^2_W}{m^2_Z\cos^2\theta_W}=1+\rho_m\label{bb5}
\eea
In the first relation the replacement of $\alpha$ with the running coupling at the Z mass $\alpha(m_Z)$ makes $\Delta r_W$
completely determined by the purely weak corrections. This relation defines $\Delta r_W$ unambigously, once the meaning of
$\alpha(m_Z)$ is specified. On the contrary, in the second relation $\Delta \rho_m$ depends on the definition of
$\sin^2\theta_W$ beyond the tree level. For LEP physics $\sin^2\theta_W$ is usually defined from the
$Z\rightarrow\mu^+\mu^-$ effective vertex. At the tree level we have:
\beq
Z\rightarrow f^+f^-=\frac{g}{2\cos\theta_W}\bar f\gamma_\mu(g^f_V-g^f_A\gamma_5)f \label{xyz}
\eeq
with $g^{f2}_A=1/4$ and $g^f_V/g^f_A=1-4|Q_f|\sin^2\theta_W$. Beyond the tree level a corrected vertex can be written down
in the same form of eq.(\ref{xyz}) in terms of modified effective couplings. Then $\sin^2\theta_W\equiv\sin^2\theta_{eff}$
is in general defined through the muon vertex:
\bea
g^\mu_V/g^\mu_A&=&1-4\sin^2\theta_{eff}\nonumber\\
\sin^2\theta_{eff}&=&(1+\Delta k)s^2_0,~~~~~~~s^2_0 c^2_0=\frac{\pi\alpha(m_Z)}{\sqrt{2}G_Fm^2_Z}\nonumber\\
g^{\mu2}_A&=&\frac{1}{4}(1+\Delta\rho)\label{bb7}
\eea
Actually, since in the SM lepton universality is only broken by masses and is in agreement with experiment within the
present accuracy, in practice the muon channel is replaced with the average over charged leptons.

We end this discussion by
writing a symbolic equation that summarises the status of what has been computed up to now for the radiative corrections
$\Delta r_W$, $\Delta \rho$ and $\Delta k$:
\beq
\Delta r_W, \Delta \rho, \Delta k=g^2 \frac{m^2_t}{m^2_W}(1+\alpha_s+\alpha^2_s) +g^2(1+\alpha_s+\sim\alpha^2_s) + g^4
\frac{m^4_t}{m^4_W} + g^4\frac{m^2_t}{m^2_W} +...\label{bb8}
\eeq
The meaning of this relation is that the one loop terms of order $g^2$ are completely known, together with their first 
order QCD corrections (the second order QCD corrections are only estimated for the $g^2$ terms not enhanced by
$m^2_t/m^2_W$), and the terms of order $g^4$ enhanced by the ratios $m^4_t/m^4_W$ or $m^2_t/m^2_W$ are also known.

In recent years new powerful tests of the SM have been performed mainly at LEP but also
at SLC and at the Tevatron. The running of LEP1 was terminated in 1995 and close-to-final results of the data
analysis are now available. The SLC is still running. The experiments at the Z resonance have enormously
improved the accuracy in the electroweak neutral current sector. The top quark has been at last
found at the Tevatron and  the errors on $m_Z$ and $\sin^2\theta_{eff}$ went down by two and one orders of magnitude
respectively since the start of LEP in 1989. The LEP2 programme is in progress. The validity
of the SM has been confirmed to a level that we can say was unexpected at the beginning. In the present data
there is no significant evidence for departures from the SM, no convincing hint of new physics (also
including the first  results from LEP2). The impressive success of the SM poses strong limitations on
the possible forms of  new physics. Favoured are models of the Higgs sector and of new physics that preserve the
SM structure  and only very delicately improve it, as is the case for fundamental Higgs(es) and
Supersymmetry. Disfavoured are models with a nearby strong non perturbative regime that  almost inevitably
would affect the radiative corrections, as for composite Higgs(es) or technicolor and its variants. 

\section{ Status of the Data}

The relevant electro-weak data together with their SM values are presented in table 1 \cite{kar},\cite{ew},\cite{ABC98}.  The
SM predictions correspond to a fit of all the available data (including the directly measured values of $m_t$
and
$m_W$) in terms of $m_t$, $m_H$ and $\alpha_s(m_Z)$, described later in sect.8, table 4.

Other important derived quantities are, for example, $N_\nu$ the number of light neutrinos,
obtained from the invisible width: $N_\nu=2.994(11)$, which shows that only three fermion generations
exist with $m_\nu <45~GeV$. This is one of the most important results of LEP. Other important quantities are the leptonic
width
$\Gamma_l$, averaged over e,
$\mu$ and
$\tau$:
$\Gamma_l= 83.90(10) MeV$ and the hadronic width $\Gamma_h= 1742.3(2.3) MeV$.   

For indicative purposes, in table  the "pulls" are also shown, defined as: pull = (data point- fit
value)/(error on data point). 
At a glance we see that the agreement with the SM is quite good. The distribution of the
pulls is statistically normal. The presence of a few $\sim2\sigma$ deviations is what is to be expected.
However it is maybe worthwhile to give a closer look at these small discrepancies.

One  persistent feature of the data is the difference between the values of
$\sin^2\theta_{eff}$ measured at LEP and at SLC (although the discrepancy is going down in the most recent data). The value of
$\sin^2\theta_{eff}$ is obtained from a set of combined asymmetries. From asymmetries one derives the ratio $x=g_V^l/g_A^l$
of the vector and axial vector couplings of the Z, averaged over the charged leptons. In turn $\sin^2\theta_{eff}$ is
defined by $x=1-4\sin^2\theta_{eff}$. SLD obtains x from the single measurement of
$A_{LR}$, the left-right asymmetry, which requires longitudinally polarized beams. The distribution of the
present measurements of
$\sin^2\theta_{eff}$ is shown in fig.4. The LEP average,
$\sin^2\theta_{eff}=0.23187(24)$, differs by
$2.2\sigma$ from the SLD value
$\sin^2\theta_{eff}=0.23101(31)$. The most
precise individual measurement at LEP is from $A^{FB}_b$: the combined LEP error on this quantity is comparable to the SLD
error, but the two values are $2.5\sigma$'s away. One might attribute this to the fact that the b measurement is more
delicate and affected by a complicated systematics. In fact one notices from
fig.4 that the value  obtained at LEP from
$A^{FB}_l$, the average for l=e, $\mu$ and $\tau$, is somewhat low (indeed quite in agreement with the SLD value). However
the statement that LEP and SLD agree on leptons while they only disagree when the b quark is considered is not quite right.
First, the low value of $\sin^2\theta_{eff}$ found at LEP from
$A^{FB}_l$  turns out to be entirely due to the $\tau$ lepton channel which leads to a central value different
than that of e and
$\mu$.  The e and $\mu$ asymmetries, which are experimentally simpler, are perfectly on top
of the SM fit. Second, if we take only e and $\mu$ asymmetries at LEP and disregard the b and $\tau$
measurements the LEP average becomes $\sin^2\theta_{eff}=0.23168(36)$, which is still $1.4\sigma$ away
from the SLD value. Thus it is difficult to find a simple explanation for the SLD-LEP discrepancy on
$\sin^2\theta_{eff}$. In the following we will tentatively use the official average
\beq
\sin^2\theta_{eff}=0.23155\pm0.00019 \label{102}
\eeq	
obtained by a simple combination of the LEP-SLC data. One could be more conservative and enlarge the error because of the
larger dispersion, but the difference would not be too large. Also, this dispersion has decreased in the most recent data. The
data-taking by the SLD experiment is still in progress and also at LEP seizable improvements on
$A_{\tau}$ and $A^{FB}_b$ are foreseen as soon as the corresponding analyses will be completed. We
hope to see the difference to be further reduced in the end.
%%%%%%%%%%%%%%%%%%%%%%%%%%%%%%%%%
\begin{figure}
\hglue 2.0cm
\epsfig{figure=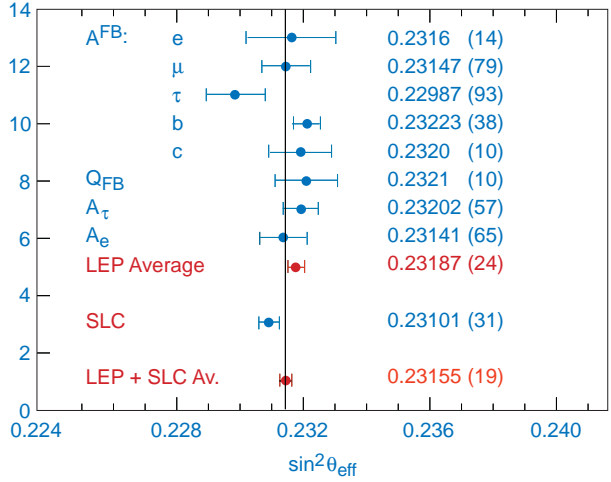, width=8cm}
\caption[]{A summary of $\sin^2\theta_{eff}$ measurements)}
\end{figure}
%%%%%%%%%%%%%%%%%%%%%%%%%%%%%%%%%%

From the above discussion one may wonder if there is evidence for something special in the $\tau$
channel, or equivalently if lepton universality is really supported by the data. Indeed this is the case: the
hint of a difference in $A^{FB}_\tau$ with respect to the corresponding e and
$\mu$ asymmetries is not confirmed by the measurements of  $A_\tau$ and  $\Gamma_\tau$ which appear normal. In principle the
fact that an anomaly shows up in $A^{FB}_\tau$  and not in
$A_\tau$ and 
$\Gamma_\tau$ is not unconceivable because the FB lepton asymmetries are very small and very precisely measured.
For example, the extraction of
$A^{FB}_\tau$ from the data on the angular distribution of $\tau$'s could be biased if the
imaginary part of the continuum was altered by some non universal new physics effect. But a more
trivial experimental problem is at the moment the most plausible option.

%%%%%%%%%%%%%%%
\begin{table}
\caption{Data on precision electroweak tests}
\vglue.3cm
\begin{center}
\footnotesize
\begin{tabular}{|l|l|l|}
\hline Quantity&Data (August'98)       & Pull\\
\hline
$m_Z$ (GeV)&91.1867(21)  &$~~0.1$\\
$\Gamma_Z$ (GeV)        &2.4939(24)  & $-0.8 $\\
$\sigma_h$ (nb) &41.491(58)     & $~~0.3$\\
$R_h$   &20.765(26)      & ~~0.7\\
$R_b$ &0.21656(74)       & ~~0.9\\
$R_c$&  0.1735(44)&    $~~0.3$ \\
$A^l_{FB}$&  0.01683(96) & $~~0.7$ \\
$A_\tau$ &      0.1431(45)       & $-0.8$ \\
$A_e$   &0.1479(51) & $~~0.25$\\
$A^b_{FB}$ &    0.0990(21)  & $-1.8$ \\
$A^c_{FB}$&     0.0709(44)      & $-0.6$\\
$A_b$ (SLD direct)   & 0.867(35) & $-1.9$\\ 
$A_c$ (SLD direct)  &  0.647(40) & $-0.5$\\ 
$\sin^2\theta_{eff}({\rm\hbox{LEP-combined}})$ & 0.23187(24) &$ ~~1.3$\\
$A_{LR}\rightarrow  \sin^2\theta_{eff}$& 0.23101(31) & $-1.8$ \\
$m_W$ (GeV) (LEP2+p$\bar p$) & 80.39(60)    & $-0.4$\\
$1-\frac{m^2_W}{m^2_Z}$ ($\nu$N) &  0.2253(21)  & $~~1.1$\\ 
$Q_W$ (Atomic PV in Cs) &  -72.11(93) & $~~1.2$\\
$m_t$ (GeV)     &173.8(5.0) & $~~0.5$\\
\hline
\end{tabular}
\end{center}
\end{table}
\vglue.3cm

A similar question can be asked for the b couplings. We have seen that the measured value of $A^{FB}_b$
is $1.8\sigma$'s below the SM fit. At the same time $R_b$ which used to show a major discrepancy is
now only about $1\sigma$'s away from the SM fit (as a result of the more sophisticated second
generation experimental techniques). 	There is a $-2.5\sigma$ deviation
on  the measured value of $A_b$ vs the SM
expectation. That  somewhat depends on how the data are combined. Let us discuss this point in detail.
$A_b$ can be measured directly at SLC by taking advantage of the beam longitudinal polarization.The SLC value (see Table 1
is  $2.2\sigma$'s below the SM value. At LEP one measures
$A^{FB}_b$    = 3/4 $A_eA_b$. One can then derive $A_b$ by inserting a value for $A_e$. The question is what
to use for $A_e$: the LEP value obtained, using lepton
universality, from the measurements of $A^{FB}_l$, $A_\tau$, $A_e$: $A_e$ = 0.1470(27), or the
combination of LEP and SLD etc.
Since we are here concerned with the b couplings it is perhaps wiser to obtain $A_b$ from
LEP by using the SM value for
$A_e$ (that is the pull-zero value of table 1):
$A^{SM}_e$   = 0.1467(15). With the value of $A_b$ derived in this way from LEP (which is  $1.7\sigma$'s below the SM
value) we finally obtain
 \beq
 A_b = 0.890\pm0.018~~~~~(\rm{LEP+SLD, A_e=A^{SM}_e: -2.5\sigma}) \label{103}
 \eeq 
In the SM $A_b$ is so close to 1
because the b quark is almost purely left-handed. $A_b$ only depends on the ratio $r=(g_R/g_L)^2$ which in
the SM is small: $r\sim 0.033$. To adequately decrease $A_b$ from its SM value one must increase r by a factor
of about 1.6, which appears large for a new physics effect. Also such a large change in
$r$ must be compensated
by decreasing $g_L^2$ by a small but fine-tuned amount in order to counterbalance the corresponding
large positive shift in $R_b$. In view of this the most likely way out is that $A^{FB}_b$ and
$A_b$ have been a bit underestimated at LEP and actually there is no anomaly in the b couplings. Then the LEP
value of $\sin^2\theta_{eff}$ would slightly move down, in the direction of decreasing the SLD-LEP discrepancy.

\section{ Precision Electroweak Data and the Standard Model}

	For the analysis of electroweak data in the SM one starts from the input parameters: some of them,
$\alpha$, $G_F$ and $m_Z$, are very well measured, some other ones, $m_{f_{light}}$, $m_t$ and
$\alpha_s(m_Z)$  are only approximately determined while $m_H$ is largely unknown. With respect to
$m_t$ the situation has much improved since the CDF/D0 direct measurement of the top quark mass. From the input parameters
one computes the radiative corrections to a sufficient precision to match the experimental capabilities. Then
one compares the theoretical predictions and the data for the numerous observables which have been measured, checks the
consistency of the theory and derives constraints on $m_t$, $\alpha_s(m_Z)$ and hopefully also on $m_H$. 

	Some comments on the least known of the input parameters are now in order.
	 The only practically
relevant terms where precise values of the light quark masses, $m_{f_{light}}$, are needed are those
related to the hadronic contribution to the photon vacuum polarisation diagrams, that
determine
$\alpha(m_Z)$. This correction is of order 6$\%$, much larger than the accuracy of a few per mille of
the precision tests. Fortunately, one can use the actual data to in principle solve the related
ambiguity. But we shall see that the left over uncertainty is still one of the main sources of
theoretical error.
As is well known  \cite{radcorr}, the QED running coupling is given by:
\begin{equation}
\alpha(s) = \frac{\alpha}{1-\Delta \alpha(s)}
\label{1a}
\end{equation}
\begin{equation}	
\Delta \alpha(s) = \Pi(s) = \Pi_\gamma(0) - {\rm Re} \Pi_\gamma(s)
\label{2a}
\end{equation}
where $\Pi(s)$ is proportional to the sum of all 1-particle irreducible vacuum
polarization diagrams. In perturbation theory $\Delta\alpha(s)$ is given by:
\begin{equation}
\Delta \alpha(s) = \frac{\alpha}{3\pi} \sum_f Q^2_f N_{Cf}\left( \log
\frac{2}{m^2_f} - \frac{5}{3} \right)
\label{3a}
\end{equation}
where $N_{Cf} = 3$ for quarks and 1 for leptons. However, the perturbative formula
is only reliable for leptons, not for quarks (because of the unknown values of the
effective quark masses). Separating the leptonic, the light quark and the top
quark contributions to $\Delta\alpha(s)$ we have:
\begin{equation}
\Delta\alpha(s) = \Delta\alpha(s)_\ell + \Delta\alpha(s)_h + \Delta\alpha(s)_t
\label{4a}
\end{equation}		
with:
\begin{equation}
\Delta\alpha(s)_\ell= 0.0331421~;~~\Delta\alpha(s)_t =
\frac{\alpha}{3\pi}~\frac{4}{15}~\frac{m^2_Z}{m^2_t} = -0.000061
\label{5a}
\end{equation}
Note that in QED there is decoupling so that the top quark contribution approaches
zero in the large $m_t$ limit. For $\Delta\alpha(s)_h$ one can use eq.(\ref{2a}) and
the Cauchy theorem to obtain the representation:
\begin{equation}
\Delta\alpha(m^2_Z)_h = -\frac{\alpha m^2_Z}{3\pi}{\rm Re}
\int^\infty_{4m^2_\pi}\frac{ds}{s}~\frac{R(s)}{s-m^2_Z-i\epsilon}
\label{6a}
\end{equation}
where $R(s)$ is the familiar ratio of the hadronic to the pointlike $\ell^+\ell^-$
cross-section from photon exchange in $e^+e^-$ annihilation. At $s$ large one can
use the perturbative expansion for $R(s)$ while at small $s$ one can use the actual
data.  In recent years there has been a lot of activity on this subject and a number of
independent new estimates of $\alpha(m_Z)$  have appeared in the literature \cite{alfaQED}. A consensus has been established
and the value used at present is
\beq
\alpha(m_Z)^{-1}=128.90\pm0.09 \label{8aa} \eeq  
As I said, for the derivation of this result th QCD theoretical prediction is actually used for large values of s where the
data do not exist. But the sensitivity of the dispersive integral to this region is strongly suppressed, so that no
important model dependence is introduced. More recently some analyses have appeared where one studied by how much the error
on $\alpha_s(m_Z)$ is reduced by using the QCD prediction down to $\sqrt{s}=m_\tau$, with the possible exception of the
regions around the charm and beauty thresholds \cite{alfanew}. These attempts were motivated by the apparent success of QCD
predictions in $\tau$ decays, despite the low $\tau$ mass (note however that the relevant currents are V-A in $\tau$ decay
but V in the present case). One finds that the central value is not much changed while the error in eq.(\ref{8aa}) is
reduced  from 0.09 down to something like 0.03-0.04, but, of course, at the price of more model dependence. For this reason,
in the following,  we shall use the more conservative value in eq.(\ref{8aa}).

	As for the strong coupling $\alpha_s(m_Z)$ the world average central value is by now quite stable. The
error is going down because the dispersion among the different measurements is much smaller in the most
recent set of data. The most important determinations of $\alpha_s(m_Z)$ are summarised in table 2 \cite{dok}. For
all entries, the main sources of error are the theoretical ambiguities which are larger than the experimental
errors. The only exception is the measurement from the electroweak precision tests, but only if one assumes
that the SM electroweak sector is correct. My personal views on the theoretical errors are reflected in
the table 2. The error on the final average is taken by all authors between
$\pm$0.003 and
$\pm$0.005 depending on how conservative one is. Thus in the following our reference value
will be \beq 
\alpha_s(m_Z) = 0.119\pm0.004 \label{111} \eeq

\begin{table}
\caption{Measurements of $\alpha_s(m_Z)$. In parenthesis we indicate if the dominant source
of errors is theoretical or experimental. For theoretical ambiguities our personal figure of merit is given.}
\vglue.3cm
\begin{center}
\footnotesize
\begin{tabular}{|l|ll|}
\hline Measurements & \multicolumn{2}{c|}{$\alpha_s(m_Z)$}\\
\hline
$R_{\tau}$ & 0.122 $\pm$ 0.006 & (Th)\\ Deep Inelastic Scattering & 0.116 $\pm$
0.005 & (Th)\\
$Y_{\rm decay}$ & 0.112 $\pm$ 0.010 & (Th)\\ Lattice QCD & 0.117 $\pm$ 0.007 &
(Th)\\
$Re^+e^-(\sqrt s < 62~{\rm GeV}$) & 0.124 $\pm$ 0.021 & (Exp)\\ Fragmentation
functions in $e^+e^-$ & 0.124 $\pm$ 0.012 & (Th)\\ Jets in $e^+e^-$ at and below
the $Z$ & 0.121 $\pm$ 0.008 & (Th)\\
$Z$ line shape (Assuming SM) & 0.120 $\pm$ 0.004 & (Exp)\\
\hline
\end{tabular}
\end{center}
\end{table}
%\vglue.3cm
	Finally a few words on the current status of the direct measurement of $m_t$. The present combined CDF/D0
result is
\beq 
m_t = 173.8\pm 5.0~GeV \label{10aa} 
\eeq
The error is so small by now that one is approaching a level
where a more careful investigation of the effects of colour rearrangement on the determination of $m_t$ is
needed. One wants to determine the top quark mass, defined as the invariant mass of its decay products (i.e.
b+W+ gluons +
$\gamma$'s). However, due to the need of colour rearrangement, the top quark and its decay products cannot be
really isolated from the rest of the event. Some smearing of the mass distribution is induced by this colour
crosstalk which involves the decay products of the top, those of the antitop and also the fragments of the
incoming (anti)protons. A reliable quantitative computation of the smearing effect on the $m_t$ 
determination is difficult because of the importance of non perturbative effects. An induced error of
the order of 1 GeV on $m_t$ could reasonably be expected. Thus further progress on the $m_t$
determination demands tackling this problem in more depth. 

	In order to appreciate the relative importance of the different sources of theoretical errors for
precision tests of the SM, we report in table 3  a comparison for the most relevant observables.	What is important to stress
is that the ambiguity from $m_t$, once by far the largest one, is by now smaller than the error from $m_H$. We also see from
table 3 that the error from
$\Delta\alpha(m_Z)$ is expecially important for $\sin^2\theta_{eff}$  and, to a lesser extent, is also sizeable for
$\Gamma_Z$ and $\epsilon_3$.  
%\vglue.3cm
\begin{table}
\caption{Errors from different sources: $\Delta^{exp}_{now}$    is
the present experimental error;
$\Delta\alpha^{-1}$ is the impact of $\Delta\alpha^{-1}=\pm0.09$;  $\Delta_{th}$
is the estimated theoretical error from higher orders; $\Delta m_t$ is from
$\Delta m_t =\pm 6 $GeV;
$\Delta m_H$ is from $\Delta m_H$ = 60--1000 GeV; $\Delta \alpha_s$ corresponds to
$\Delta \alpha_s=\pm0.003$. The epsilon parameters are defined in
sect.9.1.}
\vglue.3cm
\begin{center}
\footnotesize
\begin{tabular}{|l|l|l|l|l|l|l|}
\hline Parameter& $\Delta^{exp}_{now}$ & $\Delta \alpha^{-1}$ & $\Delta_{th}$ &
$\Delta m_t$ & $\Delta m_H$ & $\Delta \alpha_s$ \\
\hline
$\Gamma_Z$ (MeV) & $\pm$2.4 & $\pm$0.7 & $\pm$0.8 & $\pm$1.4 & $\pm$4.6 &
$\pm$1.7 \\
$\sigma_h$ (pb) & 58 & 1 & 4.3 & 3.3 & 4 & 17\\
$R_h \cdot 10^3$ & 26 & 4.3 & 5 & 2 & 13.5 & 20 \\
$\Gamma_l$ (keV) & 100 & 11 & 15 & 55 & 120 & 3.5\\
$A^l_{FB}\cdot 10^4$ & 9.6 & 4.2 & 1.3 & 3.3 & 13 & 0.18 \\
$\sin^2\theta\cdot 10^4$ & 19 & 2.3 & 0.8 & 1.9 & 7.5 & 0.1\\
$m_W$~(MeV) & 60 & 12 & 9 & 37 & 100& 2.2 \\
$R_b \cdot 10^4$ & 7.4 & 0.1 & 1 & 2.1 & 0.25 & 0\\
$\epsilon_1\cdot 10^3$ & 1.2 & & $\sim$0.1 & & & 0.2\\
$\epsilon_3\cdot 10^3$ & 1.2 & 0.5 & $\sim$0.1 & & & 0.12\\
$\epsilon_b\cdot 10^3$ & 2.1 & & $\sim$0.1 & & & 1\\
\hline
\end{tabular}
\end{center}
\end{table}
%\vglue.3cm

The most important recent advance in the theory of radiative corrections is the calculation of the 
$o(g^4m^2_t/m^2_W)$ terms in $\sin^2\theta_{eff}$, $m_W$ and, more recently in $\delta\rho$ \cite{deg}. The result implies
a small but visible correction to the predicted values but expecially a seizable decrease of the ambiguity
from scheme dependence (a typical effect of truncation). These callculations are now implemented in the fitting codes used
in the analysis of LEP data. The fitted value of the Higgs mass is lowered by about
$30~GeV$ due to this effect. 

We now discuss fitting the data in the SM. As the mass of the top quark is now rather precisely known from CDF and D0 one
must distinguish two different types of fits. In one type one wants to answer the question: is $m_t$ from radiative
corrections in agreement with the direct measurement at the Tevatron? Similarly how does $m_W$ inferred from radiative
corrections compare with the direct measurements at the Tevatron and LEP2?  For answering these interesting but somewhat
limited questions, one must clearly exclude the measurements of
$m_t$ and $m_W$ from the input set of data. Fitting all other data in terms of
$m_t$,
$m_H$ and
$\alpha_s(m_Z)$ one finds the results shown in the second column of table 4~\cite{ew}. The extracted value of $m_t$ is
typically a bit too low. For example, as shown in the table 4, from all the electroweak data except the direct
production results on $m_t$ and $m_W$,  one finds
$m_t= 158\pm 9$GeV. There is a strong correlation between $m_t$ and $m_H$.
$\sin^2\theta_{eff}$ and $m_W$ drive the fit to small values of $m_H$. Then, at small $m_H$ the widths, in particular
the leptonic width (whose prediction is nearly independent of
$\alpha_s$) drive the fit to small $m_t$. In a more general type of fit, e.g. for determining the overall consistency
of the SM or the best present estimate for some quantity, say $m_W$, one should of course not ignore the existing direct
determinations of $m_t$ and $m_W$. Then, from all the available data,  by fitting
$m_t$, $m_H$ and $\alpha_s(m_Z)$ one finds the values shown in the last column of table 4.

\begin{table}
\caption{ Standard Model fits of electroweak data.}
\vglue.3cm
\begin{center}
\footnotesize
\begin{tabular}{|l|l|l|l|}
\hline Parameter & LEP(incl.$m_W$) &All but $m_W$, $m_t$  & All Data\\
\hline
$m_t$ (GeV) & 160$+13-10$ & 158$+9-8$ & $171.3\pm4.9$\\
$m_H$ (GeV) & 66$+142-38$ & 34$+45-16$ &84$+91-51$\\
$log[m_H(GeV)]$ & 1.82$+0.50-0.37$ &  1.53$+0.37-0.28$ &  1.92$+0.32-0.41$ \\
$\alpha_s(m_Z)$ & $0.121\pm0.003$ & $0.120\pm0.003$ & $0.119\pm0.003$ \\
$\chi^2/dof$ & 4.2/9 & 14/12 & 16.4/15\\
\hline
\end{tabular}
\end{center}
\end{table} 
This is the fit also referred to in table 1. The corresponding fitted values of
$\sin^2\theta_{eff}$ and $m_W$ are: 
\beq
\sin^2\theta_{eff} =0.23156\pm0.00019\nonumber;~~~~
                        m_W = 80.370\pm0.027 ~GeV \label{10car} 
\eeq 
The fitted value of $\sin^2\theta_{eff}$ is
practically identical to the LEP+SLD average. The error of 27 MeV on
$m_W$  clearly sets up a goal for the direct measurement of $m_W$ at LEP2 and the Tevatron.

As a final
comment we want to recall that the radiative corrections are functions of $log(m_H)$. It is truly remarkable that the
fitted value of $log(m_H)$ is found to fall right into the very narrow allowed window around the value 2 specified by
the lower limit from direct searches, $m_H>\sim 90~GeV$, and the theoretical upper limit in the SM $m_H< 600-800~GeV$
(see later). Note that if the Higgs is removed from the theory, $\log m_H
\rightarrow \log \Lambda$ + constant, where $\Lambda$ is a cutoff or the scale
of the new physics that replaces the Higgs.  The control of the finite terms is
lost.  Thus the fact that from experiment, one finds $\log m_H \sim 2$ is a strong
argument in favour of the precise form of the Higgs mechanism as in the SM.  The fulfilment of this very stringent consistency check is a beautiful argument in favour of a fundamental
Higgs (or one with a compositeness scale much above the weak scale).

\section{A More General Analysis of Electroweak Data}

We now discuss an update of the epsilon analysis \cite{ABC98} which is a method to look at the data in
a more general context than the SM. The starting point is to isolate from the data that part which is due to the
purely weak radiative corrections. In fact 
the epsilon variables are defined in such a way that they are zero
in the
approximation when only effects from the SM at the tree level plus pure QED and pure QCD corrections are taken into account. This
very simple version of improved Born approximation is a good first approximation  according to the data and is independent of
$m_t$ and $m_H$. In fact the whole $m_t$ and $m_H$ dependence arises from weak loop corrections and therefore is only
contained in the epsilon variables. Thus the epsilons are extracted from the data without need of specifying
$m_t$ and $m_H$. But their predicted value in the SM or in any extension of it depend on $m_t$ and $m_H$.
This is to be compared with the competitor method based on the S, T, U variables. The latter
cannot be obtained from the data without specifying
$m_t$ and $m_H$ because they are defined as deviations from the complete SM prediction for specified $m_t$ and
$m_H$. Of course there are very many variables that vanish if pure weak loop corrections are neglected, at
least one for each relevant observable. Thus for a useful definition we choose a set of
representative observables that are used to parametrize those hot spots of the radiative corrections where
new physics effects are most likely to show up. These sensitive weak correction terms include vacuum
polarization diagrams which being potentially quadratically divergent are likely to contain all possible non
decoupling effects (like the quadratic top quark mass dependence in the SM). There are three independent
vacuum polarization contributions. In the same spirit, one must add the $Z\rightarrow b \bar b$ vertex which also
includes a large top mass dependence. Thus altogether we consider four defining observables: one asymmetry, for
example
$A_{FB}^l$, (as representative of the set of measurements that lead to the determination of
$\sin^2\theta_{eff}$), one width (the leptonic width
$\Gamma_l$ is particularly suitable because it is practically independent of $\alpha_s$), $m_W$ and $R_b$.
Here lepton universality has been taken for granted, because the data show that it is
verified within the present accuracy. The
four variables,
$\epsilon_1$, $\epsilon_2$, $\epsilon_3$ and $\epsilon_b$ are defined in one to one
correspondence with the set of observables  $A^{FB}_l$, $\Gamma_l$,
$m_W$, and $R_b$. The definition is so chosen that the quadratic top mass dependence is only
present  in
$\epsilon_1$ and
$\epsilon_b$, while the
$m_t$ dependence of
$\epsilon_2$ and $\epsilon_3$ is logarithmic. The definition of $\epsilon_1$ and $\epsilon_3$ is specified
in terms of $A^{FB}_l$ and $\Gamma_l$ only. Then adding $m_W$ or $R_b$ one obtains $\epsilon_2$ or
$\epsilon_b$. We now specify the relevant definitions in detail.

\subsection{Basic Definitions and Results}

We start from the basic observables $m_W/m_Z$, $\Gamma_l$ and  $A^{FB}_l$ and $\Gamma_b$. From these four
quantities one can isolate the corresponding dynamically significant corrections $\Delta r_W$, $\Delta \rho$, 
$\Delta k$ and $\epsilon_b$, which  contain the small effects one is trying to disentangle and are defined in
the following. First we introduce $\Delta r_W$ as obtained from $m_W/m_Z$ by the relation:
\beq
(1-\frac{m_W^2}{m_Z^2}) \frac{m_W^2}{m_Z^2}~=~\frac{\pi \alpha(m_Z)}{\sqrt{2} G_F m_Z^2 (1-\Delta r_W)}
\label{1n}
\eeq
Here $\alpha(m_Z)~=~\alpha /(1-\Delta \alpha)$ is fixed to the central value 1/128.90 so that the effect of
the running of $\alpha$ due to known physics is extracted from $1-\Delta r = (1- \Delta \alpha)(1- \Delta
r_W)$. In fact, the error on $1/\alpha(m_Z)$, as given in eq.(\ref{8aa}) would then affect $\Delta r_W$.
In order to define $\Delta
\rho$ and 
$\Delta k$ we
consider the effective vector and axial-vector couplings $g_V$ and $g_A$ of the on-shell Z to charged leptons,
given by the formulae:
\bea
\Gamma_l~&=&~\frac{G_F m^3_Z}{6\pi \sqrt{2}}(g^2_V+g_A^2) (1+\frac{3 \alpha}{4 \pi}), \nonumber \\
A_l^{FB}(\sqrt{s}&=&m_Z)~=~\frac{3g^2_Vg^2_A}{(g^2_V+g_A^2)^2}~=~\frac{3x^2}{(1+x^2)^2}. \label{2nn}
\eea
Note that $\Gamma_l$ stands for the inclusive partial width $\Gamma(Z\rightarrow l\bar l + \rm{photons})$. We
stress the following points. First, we have extracted from $(g^2_V+g_A^2)$ the factor $(1 + 3\alpha /4 \pi )$
which is induced in $\Gamma_l$ from final state radiation. Second, by the  asymmetry at the peak in
eq.(\ref{2nn}) we mean the quantity which is commonly referred to by the LEP experiments (denoted as $A^0_{FB}$
in ref.\cite{ew}), which is corrected for all QED effects, including initial and final state radiation and
also for the effect of the imaginary part of the $\gamma$ vacuum polarization  diagram. In terms of $g_A$ and
$x= g_V /g_A$, the quantities $\Delta \rho$ and 
$\Delta k$ 
are given by:
\bea
g_A~=~-\frac{\sqrt{\rho}}{2}~\sim~-\frac{1}{2}(1+\frac{\Delta \rho}{2}), \nonumber \\
x~=~\frac{g_V}{g_A}~=~1-4\sin^2\theta_{eff}~=~1-4(1+\Delta k) s_0^2.\label{3n}
\eea
Here $s_0^2$ is $\sin^2\theta_{eff}$
before non pure-QED corrections, given by:
\beq
s_0^2 c_0^2~=~ \frac{\pi \alpha(m_Z)}{\sqrt{2} G_F m_Z^2} \label{4n}
\eeq
with  $c_0^2~=~1-s_0^2$ ($s_0^2 = 0.231095$ for $m_Z~=~91.188~GeV$).
	
We now define $\epsilon_b$ from $\Gamma_b$, the inclusive partial width for $Z\rightarrow b \bar b$ according to
the relation
\beq
\Gamma_b~=~\frac{G_F m^3_Z}{6\pi \sqrt{2}}\beta (\frac{3-\beta^2}{2} g^2_{bV}~+~\beta^2 g^2_{bA}) N_C R_{QCD}
(1+\frac{\alpha}{12\pi}) \label{5n}
\eeq
where $N_C=3$ is the number of colours, $\beta=\sqrt{1-4m_b^2/m^2_Z}$, with $m_b=4.7~$
GeV, $R_{QCD}$ is the QCD correction factor given by
\beq
R_{QCD}~=~ 1~+~1.2a~-~1.1a^2~-~13a^3~;~~~a~=~\frac{\alpha_s(m_Z)}{\pi} \label{6n}
\eeq
and $g_{bV}$ and $g_{b A}$ are specified as follows
\bea
			g_{bA}~=~-\frac{1}{2}(1+\frac{\Delta \rho}{2})(1+\epsilon_b), \nonumber\\
			\frac{g_{bV}}{g_{bA}}~=~\frac{1-4/3\sin^2\theta_{eff}+\epsilon_b}{1+\epsilon_b}.\label{7n}
\eea
This is clearly not the most general deviation from the SM in the $Z\rightarrow b \bar b$ but $\epsilon_b$ is
closely related to the quantity  $-Re(\delta_{b-vertex})$ where the large
$m_t$ corrections are located in the SM.

As is well known, in the SM the quantities $\Delta r_W$, $\Delta \rho$, $\Delta k$  and $\epsilon_b$, for
sufficiently  large $m_t$, are all dominated by  quadratic terms in $m_t$ of order $G_Fm^2_t$.   As new physics
can  more easily be disentangled if not masked by large conventional $m_t$ effects, it is convenient to keep
$\Delta \rho$ and $\epsilon_b$ while trading $\Delta r_W$
and 
$\Delta k$ for two quantities with no contributions of order $G_Fm^2_t$. We thus introduce the
following linear combinations:
\bea
\epsilon_1~&=&~\Delta \rho, \nonumber \\
\epsilon_2~&=&~c^2_0 \Delta \rho~+~\frac{s^2_0 \Delta r_W}{c^2_0-s^2_0}~-~2s^2_0 \Delta k, \nonumber\\
\epsilon_3~&=&~c^2_0 \Delta \rho~+~(c^2_0-s^2_0) \Delta k. \label{8n}
\eea
The quantities $\epsilon_2$ and $\epsilon_3$ no longer contain terms of order $G_Fm^2_t$ but only logarithmic
terms in $m_t$. The leading terms for large Higgs mass, which are logarithmic, are contained in
$\epsilon_1$ and $\epsilon_3$. In the Standard Model one has the following "large"
asymptotic contributions:
\bea
\epsilon_1~&=&~\frac{3G_F m_t^2}{8 \pi^2 \sqrt{2}}~-~\frac{3G_F m_W^2}{4 \pi^2 \sqrt{2}} \tan^2{\theta_W}
\ln\frac{m_H}{m_Z}~+....,\nonumber \\
\epsilon_2~&=&~-\frac{G_F m_W^2}{2 \pi^2 \sqrt{2}}\ln\frac{m_t}{m_Z}~+....,\nonumber \\
\epsilon_ 3~&=&~\frac{G_F m_W^2}{12 \pi^2 \sqrt{2}}\ln\frac{m_H}{m_Z}~-~\frac{G_F m_W^2}{6 \pi^2
\sqrt{2}}\ln\frac{m_t}{m_Z}....,\nonumber \\
\epsilon_b~&=&~-\frac{G_F m_t^2}{4 \pi^2 \sqrt{2}}~+.... \label{9n}
\eea

The relations between the basic observables and the epsilons can be linearised, leading to the
approximate formulae
\bea
\frac{m_W^2}{m_Z^2}~&=&~\frac{m_W^2}{m_Z^2}\vert_B (1+ 1.43\epsilon_1 - 1.00\epsilon_2 - 0.86\epsilon_3),
\nonumber \\
\Gamma_l~&=&~\Gamma_l\vert_B (1+ 1.20\epsilon_1 - 0.26\epsilon_3),
\nonumber \\
A_l^{FB}~&=&~A_l^{FB} \vert_B (1+ 34.72\epsilon_1 - 45.15\epsilon_3),
\nonumber \\
\Gamma_b~&=&~\Gamma_b\vert_B (1+ 1.42\epsilon_1 - 0.54\epsilon_3 + 2.29\epsilon_b).  \label{10n}
\eea
The  Born approximations, as defined above, depend on $\alpha_s(m_Z)$ and also on $\alpha(m_Z)$. Defining
\beq
\delta \alpha_s~=~\frac{\alpha_s(m_Z)-0.119}{\pi};~~~\delta
\alpha~=~\frac{\alpha(m_Z)-\frac{1}{128.90}}{\alpha},~~~~ \label{11n}
\eeq
we have
\bea
\frac{m_W^2}{m_Z^2}\vert_B~&=&~0.768905(1-0.40\delta \alpha), \nonumber \\
\Gamma_l\vert_B~&=&~83.563(1-0.19\delta \alpha) \rm{MeV}, \nonumber \\
A_l^{FB} \vert_B~&=&~0.01696(1-34\delta \alpha), \nonumber \\
\Gamma_b\vert_B~&=&~379.8(1+1.0\delta \alpha_s-0.42\delta \alpha). \label{12nn}
\eea 
Note that the dependence on $\delta \alpha_s$ for $\Gamma_b\vert_B$, shown in eq.(\ref{12nn}), is not simply the
one loop result for $m_b=0$ but a combined effective shift which takes into account both finite mass effects
and the contribution of the known higher order terms.

%\vglue0.3cm
\begin{table} 
\caption{ Values of the epsilons in the SM as functions of $m_t$ and
$m_H$ as obtained from recent versions of ZFITTER  and TOPAZ0.
These values (in
$10^{-3}$ units) are obtained for
$\alpha_s(m_Z)$ = 0.119,
$\alpha(m_Z)$ = 1/128.90, but the theoretical predictions are essentially
independent of
$\alpha_s(m_Z)$ and $\alpha(m_Z)$.}
\begin{center}
\footnotesize
\begin{tabular}{|c|l|l|l|l|l|l|l|l|l|c|}
\hline
$m_t$ & \multicolumn{3}{|c|}{$\epsilon_1$}&\multicolumn{3}{|c|}{$\epsilon_2$}
&\multicolumn{3}{|c|}{$\epsilon_3$}&$\epsilon_b$\\ (GeV)& \multicolumn{3}{|c|}
{$m_H$ (GeV) =} &  \multicolumn{3}{|c|} {$m_H$ (GeV) =} & \multicolumn{3}{|c|}
{$m_H$ (GeV) =} & All {$m_H$}\\ & 70 & 300 & 1000 & 70 & 300 & 1000 & 70 & 300 &
1000 &\\
\hline 150      &3.55&  2.86    & 1.72 &        $-$6.85 &       $-$6.46 &       $-$5.95 &        4.98    & 6.22 &        6.81 &
$-$4.50 \\ 160 &        4.37 &  3.66 &  2.50 &  $-$7.12 &       $-$6.72 &        $-$6.20 &       4.96 &   6.18 &
6.75 &  $-$5.31
\\
 170 &  5.26 &  4.52 &  3.32 &  $-$7.43 &        $-$7.01 &        $-$6.49 &       4.94 &  6.14 &  6.69 &
$-$6.17\\
 180 &  6.19 &   5.42 &  4.18 &   $-$7.77 &       $-$7.35 &       $-$6.82 &       4.91 &  6.09 &  6.61 &
$-$7.08\\
 190 &  7.18 &   6.35 &  5.09 &  $-$8.15 &       $-$7.75 &       $-$7.20 &       4.89 &  6.03 &   6.52 &
$-$8.03\\
 200 &  8.22 &  7.34 &  6.04 &   $-$8.59 &       $-$8.18 &       $-$7.63 &       4.87 &  5.97 &  6.43 &
$-$9.01\\
\hline
\end{tabular}
\end{center}
\end{table}
%\vspace{0.3cm}

The important property of the epsilons is that, in the Standard Model, for all observables at the Z pole, the
whole dependence on $m_t$ (and $m_H$) arising from one-loop diagrams only enters through the epsilons. The same
is actually true, at the relevant level of precision, for all higher order $m_t$-dependent corrections.
Actually, the only residual $m_t$ dependence of the various observables not included in the epsilons is in the
terms of order $\alpha_s^2(m_Z)$ in the pure QCD correction factors to the hadronic widths. But this
one is quantitatively irrelevant, especially in view of the errors connected to the uncertainty on the value of
$\alpha_s(m_Z)$. The theoretical values of the epsilons in the SM from state of the art radiative corrections, also including the
recent development of ref.\cite{deg}, are given in table 5. It is important to remark that the theoretical values of the epsilons
in the SM, as given in table 5, are not affected, at the percent level or so, by reasonable variations of $\alpha_s(m_Z)$ and/or
$\alpha(m_Z)$ around their central values. By our definitions, in fact,  no terms of order $\alpha_s^n(m_Z)$ or $\alpha \ln{m_Z/m}$
contribute to the epsilons.  In terms of the epsilons, the following expressions hold, within the SM, for the
various precision observables
\bea
\Gamma_T~&=&~\Gamma_{T0}(1+1.35\epsilon_1-0.46\epsilon_3+0.35\epsilon_b), \nonumber\\
R~&=&~R_0(1+0.28\epsilon_1-0.36\epsilon_3+0.50\epsilon_b), \nonumber\\
\sigma_h~&=&~\sigma_{h0}(1-0.03\epsilon_1+0.04\epsilon_3-0.20\epsilon_b), \nonumber\\
x~&=&~x_0(1+17.6\epsilon_1-22.9\epsilon_3), \nonumber\\
R_b~&=&~R_{b0}(1-0.06\epsilon_1+0.07\epsilon_3+1.79\epsilon_b). \label{13n}
\eea
where x=$g_V/g_A$ as obtained from $A_l^{FB}$ . The quantities in
eqs.(\ref{10n}--\ref{13n}) are clearly not
independent and the redundant information is reported for convenience. By comparison with the computed radiative corrections we
obtain
\bea
\Gamma_{T0}~&=&~2489.46(1+0.73\delta \alpha_s-0.35\delta \alpha)~MeV,\nonumber \\
R_0~&=&~20.8228(1+1.05\delta \alpha_s-0.28\delta \alpha),\nonumber \\
\sigma_{h0}~&=&~41.420(1-0.41\delta \alpha_s+0.03\delta \alpha)~nb,\nonumber \\
x_0~&=&~0.075619-1.32\delta \alpha,\nonumber \\
R_{b0}~&=&~0.2182355.\label{14nn}
\eea
Note that  the quantities in eqs.(\ref{14nn}) should not be confused, at least in principle, with the
corresponding Born approximations, due to small "non universal" electroweak corrections. In practice, at the
relevant level of approximation, the difference between the two corresponding quantities is in any case
significantly smaller than the present experimental error.

In principle, any four observables could have been picked up as defining variables. 
In practice we choose those that have a more clear physical significance and are more effective in the
determination of the epsilons. In fact,  since $\Gamma_b$ is actually measured by $R_b$ (which is nearly
insensitive to $\alpha_s$), it is preferable to use directly $R_b$  itself as defining variable, as we shall do
hereafter. In practice, since the value in eq.(\ref{14nn}) is practically indistinguishable from the Born
approximation of $R_b$, this determines no change in any of the equations given above but simply requires the
corresponding replacement among the defining relations of the epsilons.

\subsection{Experimental Determination of the Epsilon Variables}

The values of the epsilons as obtained, following the
specifications in the previous sect.9.1, from the defining variables $m_W$, $\Gamma_l$, $A^{FB}_l$ and $R_b$
are shown in the first column of table 6.

\begin{table} 
\caption{Experimental values of the epsilons in the SM from different sets of data.
These values (in $10^{-3}$ units) are obtained for
$\alpha_s(m_Z) = 0.119\pm0.003$,
$\alpha(m_Z) = 1/128.90\pm0.09$, the corresponding uncertainties being included in the quoted errors.}
\vglue.3cm
\begin{center}
\footnotesize
\begin{tabular}{|l|l|l|l|l|}
\hline $\epsilon~~~10^3$  &Only def. quantities &All asymmetries &All High Energy & All Data\\
\hline
$\epsilon_1~10^3$ &$4.1\pm1.2$ &$4.3\pm1.2$ &$4.0\pm1.1$  &$3.7\pm1.1$ \\
$\epsilon_2~10^3$ &$-8.2\pm2.1$ &$-8.8\pm1.9$ &$-9.0\pm2.0$  &$-9.3\pm2.0$ \\
$\epsilon_3~10^3$ &$3.3\pm1.9$ &$4.4\pm1.2$ &$4.2\pm1.1$  &$3.9\pm1.1$ \\
$\epsilon_b~10^3$ &$-4.3\pm1.9$ &$-4.4\pm1.9$ &$-4.8\pm1.9$  &$-4.6\pm1.9$  \\
\hline
\end{tabular}
\end{center}
\end{table}

\vglue.3cm
To proceed further and include other measured observables in the analysis we need to make some
dynamical assumptions. The minimum amount  of model dependence is introduced by including other purely
leptonic quantities at the Z pole such as $A_{\tau}$, $A_e$ (measured  from the angular
dependence of the $\tau$ polarization) and $A_{LR}$ (measured by SLD). For this step, one is simply
assuming that the different leptonic asymmetries are equivalent measurements of $\sin^2\theta_{eff}$. We add, as usual, 
the measure of
$A^{FB}_b$ because this observable is dominantly sensitive to the leptonic vertex. We then use the combined value
of $\sin^2\theta_{eff}$ obtained from the whole set of asymmetries measured at LEP and SLC given in
eq.(\ref{102}). At this stage the
best values of the epsilons are shown in the second column of table 6. In figs. 5-8  we report the 1$\sigma$ ellipses in the
indicated
$\epsilon_i$-$\epsilon_j$ planes that correspond to this set of
input data.
%%%%%%%%%%%%%%%%%%%%%%%%%%%%%%%%%
\begin{figure}
\hglue 2.0cm
\epsfig{figure=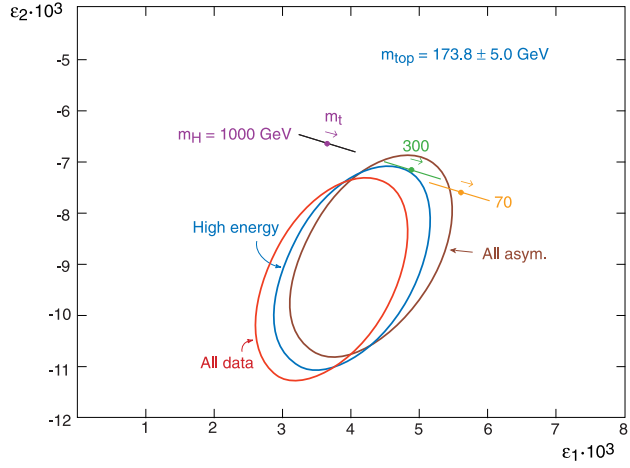, width=8.2cm}
\caption[]{Data vs theory in the $\epsilon_2$-$\epsilon_1$ plane. The origin point corresponds to the "Born"
approximation obtained from the SM at tree level plus pure QED and pure QCD corrections. The predictions of the
full SM (also including the improvements of ref.\cite{deg}) are shown for $m_H$ = 70, 300 and 1000 GeV and
$m_t=175.6\pm5.5~GeV$ (a segment for each $m_H$ with the arrow showing the direction of 
$m_t$ increasing from
$-1\sigma$ to $+1\sigma$). The three
$1-\sigma$ ellipses ($38\%$ probability contours) are obtained from a) "All Asymm." :$\Gamma_l$, $m_W$ and
$\sin^2\theta_{eff}$ as obtained from the combined asymmetries (the value in
eq. (\ref{102})); b) "All High
En.": the same as in a) plus all the hadronic variables at the Z; c) "All Data": the same as in b) plus the low
energy data.}
\end{figure}
%%%%%%%%%%%%%%%%%%%%%%%%%%%%%%%%%%
 %%%%%%%%%%%%%%%%%%%%%%%%%%%%%%%%%
\begin{figure}
\hglue 2.0cm
\epsfig{figure=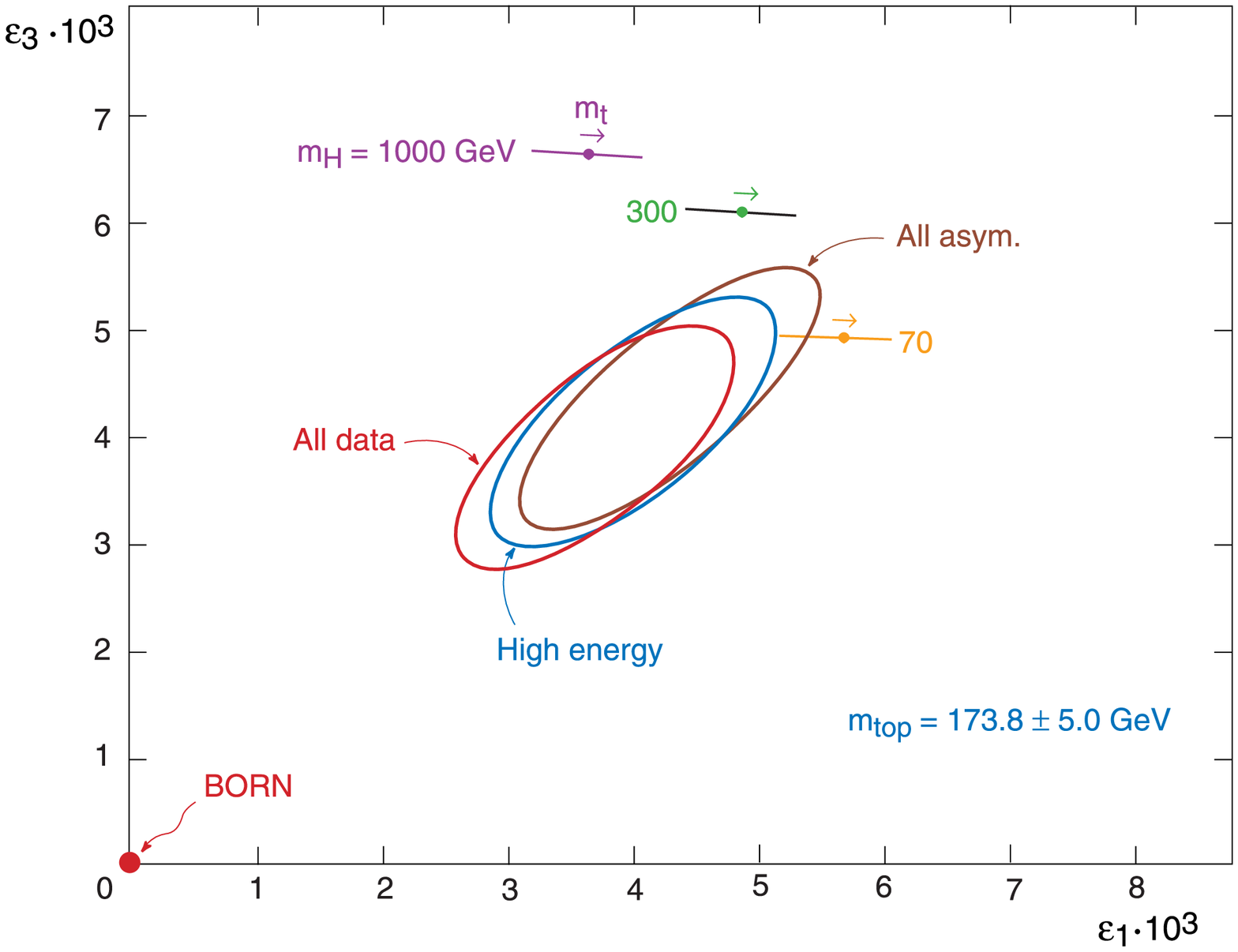, width=8.2cm}
\caption[]{Data vs theory in the $\epsilon_3$-$\epsilon_1$ plane (notations as in fig.5)}
\end{figure}
%%%%%%%%%%%%%%%%%%%%%%%%%%%%%%%%%%
%%%%%%%%%%%%%%%%%%%%%%%%%%%%%%%%%
\begin{figure}
\hglue 2.0cm
\epsfig{figure=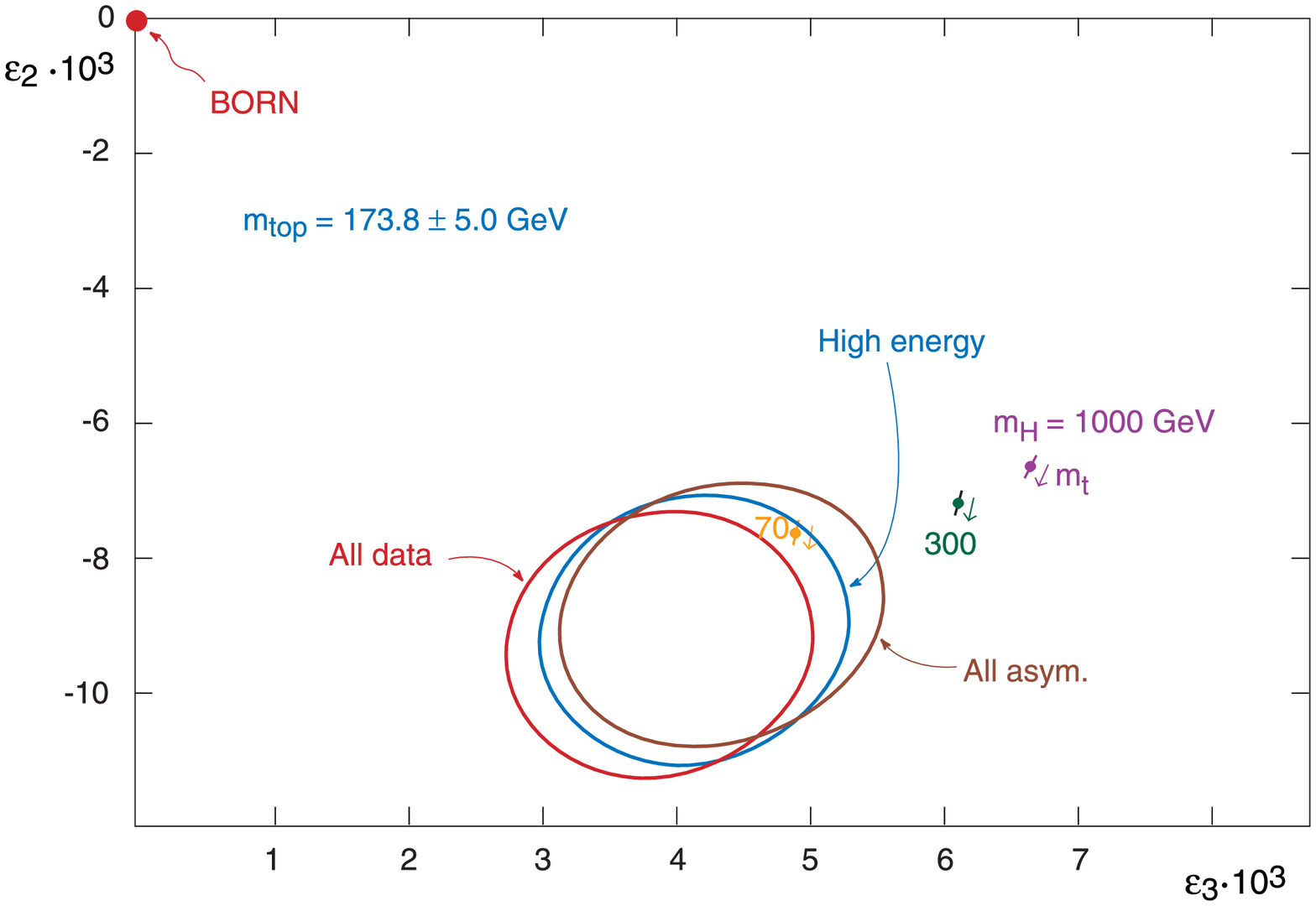, width=9.0cm}
\caption[]{Data vs theory in the $\epsilon_2$-$\epsilon_3$ plane (notations as in fig.5)}
\end{figure}
%%%%%%%%%%%%%%%%%%%%%%%%%%%%%%%%%%
%%%%%%%%%%%%%%%%%%%%%%%%%%%%%%%%%
\begin{figure}
\hglue 2.0cm
\epsfig{figure=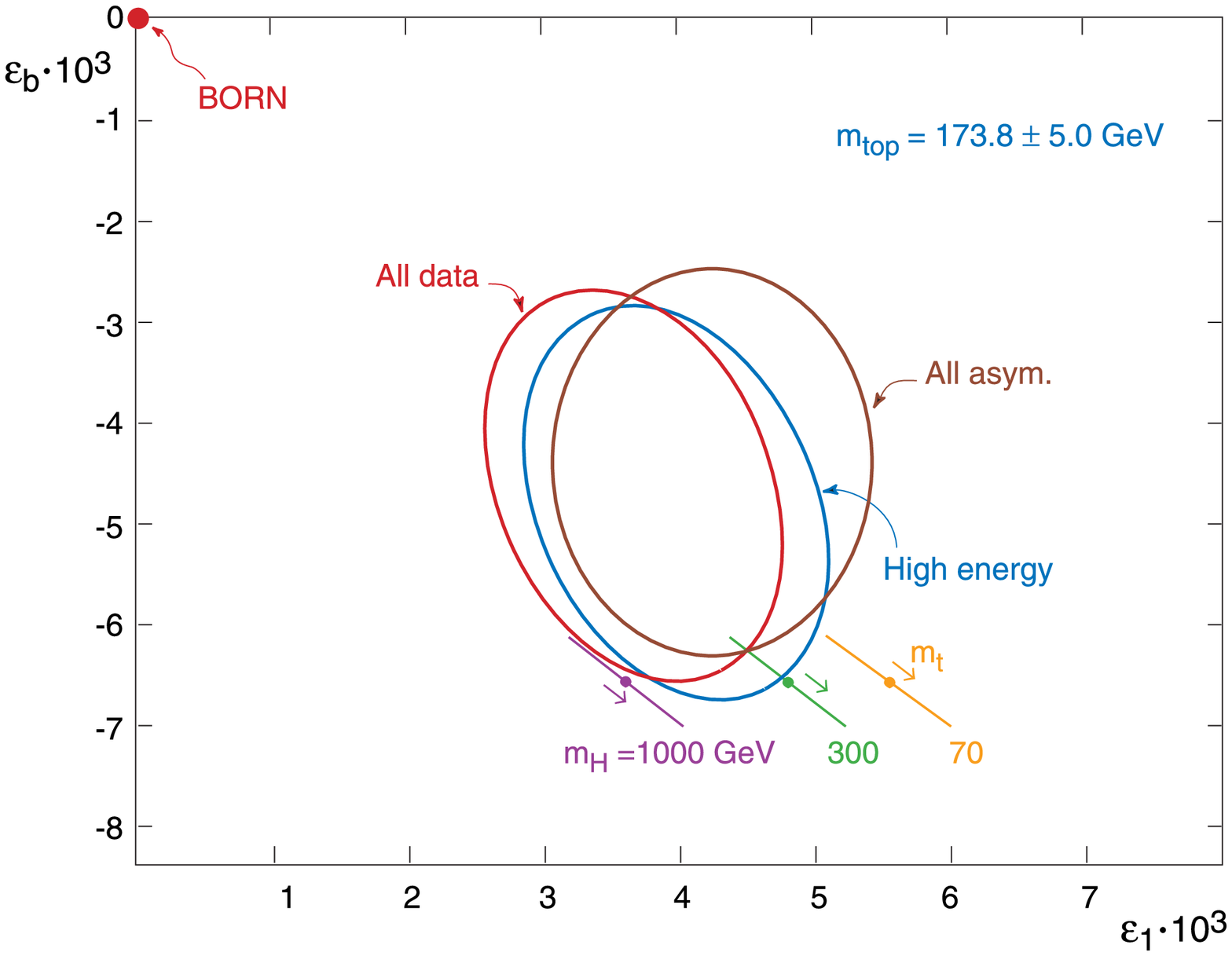, width=9.0cm}
\caption[]{Data vs theory in the $\epsilon_b$-$\epsilon_1$ plane (notations as in fig.5)}
\end{figure}
%%%%%%%%%%%%%%%%%%%%%%%%%%%%%%%%%%

	All observables measured on the Z peak at LEP can be included in the analysis provided that we assume
that all deviations from the SM are only contained in vacuum polarization diagrams (without demanding
a truncation of the $q^2$ dependence of the corresponding functions) and/or the $Z\rightarrow b\bar
b$  vertex. From a global fit of the data on $m_W$,  $\Gamma_T$,  $R_h$, $\sigma_h$,  $R_b$ and
$\sin^2\theta_{eff}$ (for LEP data, we have taken the correlation matrix for $\Gamma_T$,  $R_h$ and
$\sigma_h$ given by the LEP experiments \cite{ew}, while we have considered the additional information
on $R_b$ and $\sin^2\theta_{eff}$  as independent) we obtain the values shown in the third column of table
6. The comparison of theory and experiment at this stage is also shown in figs. 5-8. More detailed
information is shown in fig. 9, which refers to the level when also hadronic data are taken
into account. But in fig.9 we compare the results obtained if $\sin^2\theta_{eff}$ is extracted in turn from
different asymmetries among those listed in fig.4. The ellipse marked "average" is the same as the one
labelled "All high en." in fig.6 and corresponds to the value of $\sin^2\theta_{eff}$ which is shown on the
figure (and in eq.(\ref{102})). We confirm that the value from $A_{LR}$ is far away from the SM given the
experimental value of $m_t$ and the bounds on $m_H$ and would correspond to very small values of
$\epsilon_3$ and of $\epsilon_1$. We see also that while the $\tau$ FB asymmetry is also on the low side,
the combined e and $\mu$ FB asymmetry are right on top of the average. Finally the b FB asymmetry is on the
high side.
%%%%%%%%%%%%%%%%%%%%%%%%%%%%%%%%%
\begin{figure}
\hglue 1.0cm
\epsfig{figure=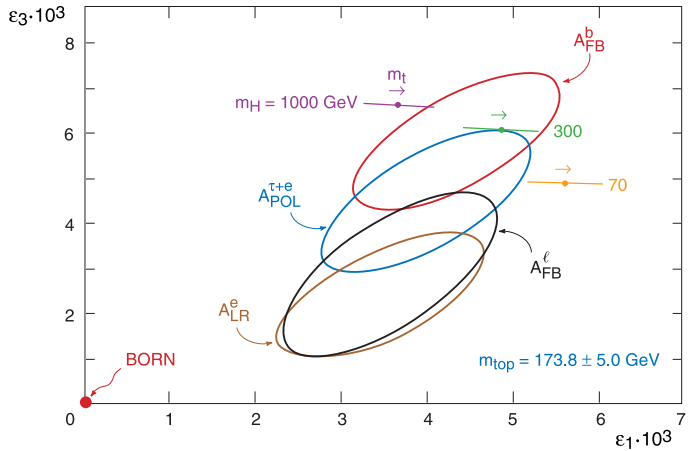, width=9cm}
\caption[]{Data vs theory in the $\epsilon_3$-$\epsilon_1$ plane (notations as in fig.5). The ellipse indicated with "Average"
corresponds to the case "All high en" of fig.6 and is obtained from the combined value of $sin^2\theta_{eff}$.  The other ellipses
are obtained by replacing the combined $sin^2\theta_{eff}$ with the values obtained in turn from each individual asymmetry as shown
by the labels.}
\end{figure}
%%%%%%%%%%%%%%%%%%%%%%%%%%%%%%%%%%

	 To include in our analysis lower energy observables as well, a stronger hypothesis needs to be
made: vacuum polarization diagrams are allowed to vary from the SM  only in their constant and first
derivative terms in a $q^2$ expansion. In such a case, one can, for example, add to the
analysis the ratio
$R_\nu$ of neutral to charged current processes in deep inelastic neutrino scattering on
nuclei\cite{33},\cite{nutev}
 the "weak charge" $Q_W$  measured in atomic parity violation experiments on Cs \cite{34} 
and the measurement of $g_V/g_A$ from $\nu_\mu e$ scattering \cite{35}. In this way one obtains  the
global fit given in the fourth column of table 6 and shown in figs. 5-8. In fig. 10 we see the ellipse in the $\epsilon
_1$-$\epsilon _3$ plane that is obtained from the low energy data by themselves. It is interesting that the tendency
towards low values of $\epsilon
_1$ and $\epsilon _3$ is present in the low energy data as in the high energy ones. Note that the low energy data
by themselves are actually compatible with the "Born" approximation. With the progress of LEP the low energy
data, while important as a check that no deviations from the expected
$q^2$ dependence arise, play a lesser role in the global fit. This does not mean that they are not important. For
example, the measured parity violation in atomic physics provides the best limits on possible new physics in the
electron-quark sector. When HERA suggested the presence of leptoquarks, the limits from atomic parity violation
practically excluded all possible parity violating four fermiom electron-quark contact terms. So low energy data
are no more powerful enough to improve the determination of the parameters if the SM is assumed, but they are a
very powerful constraint on new physics models. The best values of the $\epsilon$'s from all the data are at present:
\bea
\epsilon_1~10^3&=& 3.7\pm1.1 \nonumber\\
\epsilon_2~10^3&=& -9.3\pm2.0\nonumber\\
\epsilon_3~10^3&=& 3.9\pm1.1 \nonumber\\
\epsilon_b~10^3&=& -4.6\pm1.9. \label{epsf}
\eea
%%%%%%%%%%%%%%%%%%%%%%%%%%%%%%%%%
\begin{figure}
\hglue 2.0cm
\epsfig{figure=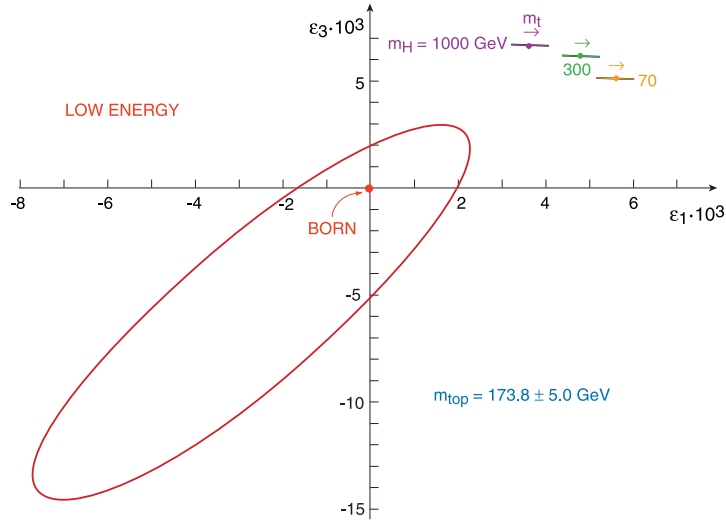, width=9.5cm}
\caption[]{Data vs theory in the $\epsilon_3$-$\epsilon_1$ plane (notations as in fig.5). Here the ellipse from the low energy data by
themselves is plotted (deep inelastic neutrino scattering, atomic parity violation and $\nu_\mu-e$ scattering.
}
\end{figure}
%%%%%%%%%%%%%%%%%%%%%%%%%%%%%%%%%%
Note that the present ambiguity on the value of
$\delta\alpha^{-1}(m_Z) =\pm0.09$ corresponds to an uncertainty on
$\epsilon_3$ (the other epsilons are not much affected) given by $\Delta\epsilon_3~10^3 =\pm0.6$. Thus the
theoretical error is still confortably less than the experimental error. In fig.11 we present a summary of the
experimental values of the epsilons as compared to the SM predictions as functions of $m_t$ and $m_H$, which
shows agreement within $1\sigma$, but the central value of $\epsilon_1$, $\epsilon_2$ and $\epsilon_3$ are all
low, while the central value of $\epsilon_b$ is shifted upward with respect to the SM as a
consequence of the still imperfect matching of $R_b$.
%%%%%%%%%%%%%%%%%%%%%%%%%%%%%%%%%
\begin{figure}
\hglue 1.5cm
\epsfig{figure=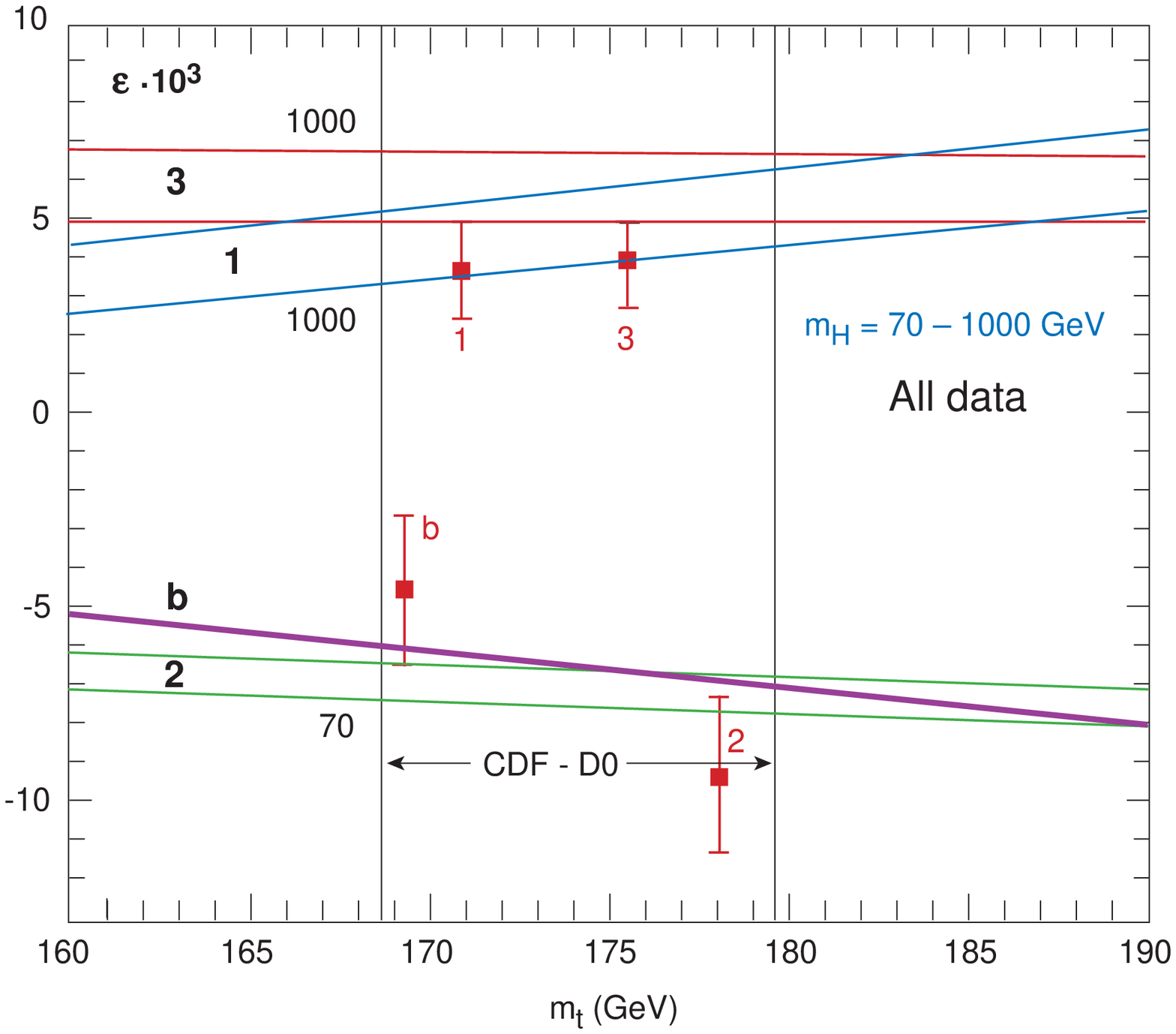, width=9cm}
\caption[]{The bands (labeled by the $\epsilon$ index) are the predicted values of the epsilons in the SM as functions of
$m_t$ for
$m_H~=~70-1000$ GeV (the $m_H$ value corresponding to one edge of the band is indicated). The CDF/D0 experimental
1-$\sigma$ range of $m_t$ is shown. The esperimental results for the epsilons from all data are displayed (from the last
column of table 6). The position of the data on the $m_t$ axis has been arbitrarily chosen and has no particular
meaning.
}
\end{figure}
%%%%%%%%%%%%%%%%%%%%%%%%%%%%%%%%%%
A number of
interesting features are clearly visible from figs.5-11. First, the good agreement with the SM and the
evidence for weak corrections, measured by the distance of the data from the improved Born approximation point
(based on tree level SM plus pure QED or QCD corrections). There is by now a solid evidence for departures from
the improved Born approximation where all the epsilons vanish. In other words a clear evidence for the pure
weak radiative corrections has been obtained and LEP/SLC are now measuring the various components of these
radiative corrections. For example, some authors \cite{39} have studied the sensitivity of the data to a
particularly interesting subset of the weak radiative corrections, i.e. the purely bosonic part. These terms
arise from virtual exchange of gauge bosons and Higgses. The result is that indeed the measurements are
sufficiently precise to require the presence of these contributions in order to fit the data. Second, the
general results of the SM fits are reobtained from a different perspective. We see the preference for light
Higgs manifested by the tendency for
$\epsilon_3$ to be rather on the low side. Since $\epsilon_3$ is practically independent of $m_t$, its low value
demands $m_H$ small. If the Higgs is light then the preferred value of
$m_t$ is somewhat lower than the Tevatron result (which in the epsilon analysis is not included among the input
data). This is because also the value of $\epsilon_1\equiv \delta \rho$, which is determined by the widths, in
particular by the leptonic width, is somewhat low. In fact
$\epsilon_1$ increases with $m_t$ and, at fixed $m_t$, decreases with $m_H$, so that for small $m_H$ the low
central value of $\epsilon_1$ pushes $m_t$ down. Note that also the central value of $\epsilon_2$ is on
the low side, because the experimental value of $m_W$ is a little bit too large. Finally, we see that adding the
hadronic quantities or the low energy observables hardly makes a difference in the
$\epsilon_i$-$\epsilon_j$ plots with respect to the case with only the leptonic variables being included (the
ellipse denoted by "All Asymm."). But, for example for the
$\epsilon_1$-$\epsilon_3$ plot, while the leptonic ellipse contains the same information as one could obtain from a
$\sin^2\theta_{eff}$ vs $\Gamma_l$ plot, the content of the other two ellipses is much larger because it
shows that the hadronic as well as the low energy quantities match the leptonic variables without need of any new
physics. Note that the experimental values of $\epsilon_1$ and
$\epsilon_3$ when the hadronic quantities are included also depend on the input value of $\alpha_s$ given
in eq.(\ref{111}).

The good agreement of the fitted epsilon values with the SM impose strong constraints on possible forms of new physics.
Consider, for example, new quarks or leptons. Mass splitted multiplets contribute to $\Delta\epsilon_1$, in analogy to
the t-b quark doublet. Recall that $\Delta\epsilon_1\sim+9.5~10^{-3} $ for the t-b doublet, which is about eight
$\sigma$'s in terms of the present error~\cite{bama}. Even mass degenerate multiplets are strongly constrained. They
contribute to $\Delta\epsilon_3$ according to~\cite{besi}
\beq
\Delta\epsilon_3 \sim N_C \frac{G_Fm_W^2}{8\pi^2\sqrt{2}}\frac{4}{3}(T_{3L}-T_{3R})^2 \label{129}\\
\eeq For example a new left-handed quark doublet, degenerate in mass, would contribute $\Delta\epsilon_3\sim +1.3~
10^{-3}$, that is about one $\sigma$, but in the wrong direction, in the sense that the experimental value of
$\epsilon_3$ favours a displacement, if any, with negative sign. Only vector fermions $(T_{3L}=T_{3R})$ are not
constrained. In particular, naive technicolour models , that introduce several new technifermions, are strongly
disfavoured because they tend to produce large corrections with the wrong sign to $\epsilon_1$,
$\epsilon_3$ and also to $\epsilon_b$ \cite{chiv}. 

\section{ Why Beyond the Standard Model?}

	Given the striking success of the SM why are we not satisfied with that theory? Why not just find the Higgs particle,
for completeness, and declare that particle physics is closed? The main reason is that there are strong conceptual
indications for physics beyond the SM.  There are also some phenomenological hints. 

\subsection{Conceptual Problems with the Standard Model}

	It is considered highly unplausible that the origin of the electro-weak symmetry breaking can be explained by the
standard Higgs mechanism, without accompanying new phenomena. New physics should be manifest at energies in the TeV
domain. This conclusion follows fron an extrapolation of the SM at very high energies. The computed behaviour of the
$SU(3)\otimes SU(2)\otimes U(1)$ couplings with energy clearly points towards the unification of the electro-weak and
strong forces (Grand Unified Theories: GUTs) at scales of energy
$M_{GUT}\sim  10^{14}-10^{16}~ GeV$ which are close to the scale of quantum gravity, $M_{Pl}\sim 10^{19}~ GeV$
\cite{ross}.  One can also imagine  a unified theory of all interactions also including gravity (at present superstrings provide
the best attempt at such a theory). Thus GUTs and the realm of quantum gravity set a very distant energy horizon that modern
particle theory cannot anymore ignore. Can the SM without new physics be valid up to such large energies? This appears unlikely
because the structure of the SM could not naturally explain the relative smallness of the weak scale of mass, set by the Higgs
mechanism at $m\sim 1/\sqrt{G_F}\sim  250~ GeV$  with $G_F$ being the Fermi coupling constant. The weak scale m is $\sim 10^{17}$
times smaller than 
$M_{Pl}$. Even if the weak scale is set near 250~ GeV at the classical level, quantum fluctuations would naturally
shift  it up to where new physics starts to apply, in particular up to  $M_{Pl}$ if there was no new physics up to
gravity. This so-called hierarchy problem \cite{ssi} is related to the presence of fundamental scalar fields in the
theory with quadratic mass divergences and no protective extra symmetry at m=0. For fermions, first, the divergences
are logaritmic and, second, at m=0 an additional symmetry, i.e. chiral  symmetry, is restored. Here, when talking of
divergences we are not worried of actual infinities. The theory is renormalisable and finite once the dependence on the
cut off is absorbed in a redefinition of masses and couplings. Rather the hierarchy problem is one of naturalness. If
we consider the cut off as a manifestation of new physics that will modify the theory at large energy scales, then it
is relevant to look at the dependence of physical quantities on the cut off and to demand that no unexplained
enormously accurate cancellation arise. 

	According to the above argument the observed value of $m\sim 250~ GeV$ is indicative of the existence of new physics
nearby. There are two main possibilities. Either there exist fundamental scalar Higgses but the theory is stabilised by
supersymmetry, the boson-fermion symmetry that would downgrade the degree of divergence from quadratic to logarithmic.
For approximate supersymmetry the cut off is replaced by the splitting between the normal particles and their
supersymmetric partners. Then naturalness demands that this splitting (times the size of the weak gauge coupling) is of
the order of the weak scale of mass, i.e. the separation within supermultiplets should be of the order of no more than
a few TeV. In this case the masses of most supersymmetric partners of the known particles, a very large managerie of
states, would fall, at least in part, in the discovery reach of the LHC. There are consistent, fully formulated field
theories constructed on the basis of this idea, the simplest one being the MSSM \cite{43}. Note that all normal
observed states are those whose masses are forbidden in the limit of exact
$SU(2)\otimes U(1)$. Instead for all SUSY partners the masses are allowed in that limit. Thus when supersymmetry is
broken in the TeV range but $SU(2)\otimes U(1)$ is intact only s-partners take mass while all normal particles remain
massless. Only at the lower weak scale the masses of ordinary particles are generated. Thus a simple criterium exists
to understand the difference between particles and s-particles.

	The other main avenue is compositeness of some sort. The Higgs boson is not elementary but either a bound state of
fermions or a condensate, due to a new strong force, much stronger than the usual strong interactions, responsible for
the attraction. A plethora of new "hadrons", bound by the new strong force would  exist in the LHC range. A serious
problem for this idea is that nobody sofar has been  able to build up a realistic model along these lines, but that
could eventually be explained by a lack of ingenuity on the theorists side. The most appealing examples are technicolor
theories \cite{chiv}. These models where inspired by the breaking of chiral symmetry in massless QCD induced by
quark condensates. In the case of the electroweak breaking new heavy techniquarks must be introduced and the scale
analogous to $\Lambda_{QCD}$ must be about three orders of magnitude larger. The presence of such a large force
relatively nearby has a strong tendency to clash with the results of the electroweak precision tests. Another
interesting idea is to replace the Higgs by a
$t\bar t$ condensate \cite{uui}. The Yukawa coupling of the Higgs to the $t\bar t$ pair becomes a four fermion 
$\bar tt\bar tt$  coupling with the corresponding strenght. The strong force is in this case provided by the large top
mass. At first sight this idea looks great:  no fundamental scalars, no new states. But, looking closely, the
advantages are largely illusory. First, in the SM the required value of $m_t$ is too large $m_t\geq 220 ~GeV$ or so.
Also a tremendous fine tuning is required, because $m_t$ would naturally be of the order of $M_{GUT}$ or $M_{Pl}$ if no
new physics is present (the hierarchy problem in a different form!). Supersymmetry could come to the rescue in this
case also. In a minimal SUSY version the required value of the top mass is lowered \cite{vvi},  $m_t\sim 205
\sin{\beta}$~ GeV. But the resulting theory is physically indistinguishable from the MSSM with small $\tan{\beta}$, at
least at low energies \cite{wwi}. This is because a strongly coupled Higgs looks the same as a $t\bar t$ pair.

	The hierarchy problem is certainly not the only conceptual problem of the SM. There are many more: the proliferation
of parameters, the mysterious pattern of fermion masses and so on. But while most of these problems can be postponed to
the final theory that will take over at very large energies, of order $M_{GUT}$ or
$M_{Pl}$, the hierarchy problem arises from the unstability of the low energy theory and requires a solution at
relatively low energies. A supersymmetric extension of the SM provides a way out which is well defined, computable and
that preserves all virtues of the SM. The necessary SUSY breaking can be introduced through soft terms that do not
spoil the stability of scalar masses. Precisely those terms arise from supergravity when it is spontaneously broken in
a hidden sector \cite{yyi}. But alternative mechanisms of SUSY breaking are also being considered
\cite{gauge}. In the most familiar approach SUSY is broken in a hidden sector and the scale of SUSY breaking is very
large of order
$\Lambda\sim\sqrt{G^{-1/2}_F M_P}$  where
$M_P$ is the Planck mass. But since the hidden sector only communicates with the visible sector through gravitational
interactions the splitting of the SUSY multiplets is much smaller, in the TeV energy domain, and the Goldstino is
practically decoupled. In an alternative scenario the (not so much) hidden sector is connected to the visible one by
ordinary gauge interactions. As these are much stronger than the gravitational interactions, $\Lambda$ can be much
smaller, as low as 10-100 TeV. It follows that the Goldstino is very light in these models (with mass of order or below
1 eV typically) and is the lightest, stable SUSY particle, but its couplings are observably large. The radiative decay
of the lightest neutralino into the Goldstino leads to detectable photons. The signature of photons comes out naturally
in this SUSY breaking pattern: with respect to the MSSM, in the gauge mediated model there are typically more photons
and less missing energy. Gravitational and gauge mediation are extreme alternatives: a spectrum of intermediate cases
is conceivable. The main appeal of gauge mediated models is a better protection against flavour changing neutral
currents. In the gravitational version even if we accept that gravity leads to degenerate scalar masses at a scale near
$M_{Pl}$ the running of the masses down to the weak scale can generate mixing induced by the large masses of the third
generation fermions \cite{ane}.

 \subsection{Hints from Experiment}

\subsubsection{Unification of Couplings}

At present the most direct
phenomenological evidence in favour of supersymmetry is obtained from the unification of couplings in GUTs.
Precise LEP data on $\alpha_s(m_Z)$ and $\sin^2{\theta_W}$ confirm what was already known with less accuracy:
standard one-scale GUTs fail in predicting $\sin^2{\theta_W}$ given
$\alpha_s(m_Z)$ (and $\alpha(m_Z)$) while SUSY GUTS \cite{zzi} are in agreement with the present, very precise,
experimental results. According to the recent analysis of ref.~\cite{aaii}, if one starts from the known values of
$\sin^2{\theta_W}$ and $\alpha(m_Z)$, one finds for $\alpha_s(m_Z)$ the results:
\bea
		\alpha_s(m_Z) = 0.073\pm 0.002 ~~~~~      	(\rm{Standard~ GUTS})\nonumber \\	
		\alpha_s(m_Z) = 0.129\pm0.010~~~~~  (\rm{SUSY~ GUTS})
\label{130}
\eea
to be compared with the world average experimental value $\alpha_s(m_Z)$ =0.119(4).

\subsubsection{Dark Matter}

There is solid astrophysical and cosmological evidence \cite{kol}, \cite{spi} that most of the matter in the universe
does not emit electromagnetic radiation, hence is "dark". Some of the dark matter must be baryonic but most of it must
be non baryonic. Non baryonic dark matter can be cold or hot. Cold means non relativistic at freeze out, while hot is
relativistic. There is general consensus that most of the non baryonic dark matter must be cold dark matter. A couple
of years ago the most likely composition was quoted to be around $80\%$ cold and $20\%$ hot. At present it appears to me
that the need of a sizeable hot dark matter component is more uncertain. In fact, recent experiments have indicated the
presence of a previously disfavoured cosmological constant component in
$\Omega =\Omega_m+\Omega_{\Lambda}$ \cite{kol}. Here
$\Omega$ is the total matter-energy density in units of the critical density, $\Omega_m$ is the matter component
(dominated by cold dark matter) and $\Omega_{\Lambda}$ is the cosmological component. Inflationary theories almost
inevitably predict
$\Omega=1$ which is consistent with present data. At present, still within large uncertainties, the approximate
composition is indicated to be
$\Omega_m\sim 0.4$ and
$\Omega_{\Lambda}\sim0.6$ (baryonic dark matter gives $\Omega_b\sim0.05$). 

The implications for particle physics is that certainly there must exist a source of cold dark matter. By far the
most appealing candidate is the neutralino, the lowest supersymmetric particle, in general a superposition of
photino, Z-ino and higgsinos. This is stable in supersymmetric models with R parity conservation, which are the
most standard variety for this class of models (including the Minimal Supersymmetric Standard Model:MSSM). A
neutralino with mass of order 100 GeV would fit perfectly as a cold dark matter candidate. Another common
candidate for cold dark matter is the axion, the elusive particle associated to a possible solution of the strong
CP problem along the line of a spontaneously broken Peccei-Quinn symmetry. To my knowledge and taste this option is
less plausible than the neutralino. One favours supersymmetry for very diverse conceptual and
phenomenological reasons, as described in the previous sections, so that neutralinos are sort of standard by now.
For hot dark matter, the self imposing candidates are neutrinos. If we demand a density fraction
$\Omega_{\nu}\sim0.1$ from neutrinos, then it turns out that the sum of stable neutrino masses should be around 5
eV. 

\subsubsection{Baryogenesis}
 Baryogenesis is interesting because it could occur at the weak
scale \cite{rub} but not in the SM. For baryogenesis one needs the three famous Sakharov conditions~\cite{sak}: B
violation, CP violation and no termal equilibrium. In principle these conditions could be verified in the SM. B is
violated by instantons when kT is of the order of the weak scale (but B-L is conserved). CP is violated by the CKM
phase and out of equilibrium conditions could be verified during the electroweak phase transition. So the
conditions for baryogenesis appear superficially to be present for it to occur at the weak scale in the SM.
However, a more quantitative analysis \cite{rev}, \cite{cw1} shows that baryogenesis is not possible
in the SM because there is not enough CP violation and the phase transition is not sufficiently strong first order,
unless
$m_H<80~GeV$, which is by now excluded by LEP. Certainly baryogenesis could also occur  below the GUT scale, after
inflation. But only that part with
$|B-L|>0$ would survive and not be erased at the weak scale by instanton effects. Thus baryogenesis at $kT\sim
10^{12}-10^{15}~GeV$ needs B-L violation at some stage like for $m_\nu$. The two effects could be related if
baryogenesis arises from leptogenesis \cite{lg} then converted into baryogenesis by instantons. While baryogenesis
at a large energy scale is thus not excluded it is interesting that recent studies have shown that baryogenesis at
the weak scale could be possible in the MSSM \cite{cw1}. In fact, in this model there are additional sources of CP
violations and the bound on $m_h$ is modified by a sufficient amount by the presence of scalars with large
couplings to the Higgs sector, typically the s-top. What is required is that $m_h\sim 80-110~GeV$ (in the
LEP2 range!), a s-top not heavier than the top quark and, preferentially, a small $\tan{\beta}$.

\subsubsection{Neutrino Masses}

Recent data from Superkamiokande \cite{SK}(and also MACRO \cite{MA}) have provided a more
solid experimental basis for neutrino oscillations as an explanation of the atmospheric neutrino
anomaly. In addition the solar neutrino deficit is also probably an indication of a different
sort of neutrino oscillations. Results from the laboratory experiment by the LNSD
collaboration \cite{LNSD} can also be considered as a possible indication of yet another type
of neutrino oscillation. But the preliminary data from Karmen \cite{KA} have failed to
reproduce this evidence. The case of LNSD oscillations is far from closed but one can
tentatively assume, pending the results of continuing experiments, that the signal will not
persist. Then solar and atmospheric neutrino oscillations can possibly be explained in terms
of the three known flavours of neutrinos without invoking extra sterile species. Neutrino
oscillations for atmospheric neutrinos require
$\nu_{\mu}\rightarrow\nu_{\tau}$ with $\Delta m^2_{atm}\sim 2~10^{-3}~eV^2$ and a nearly
maximal mixing angle
$\sin^2{2\theta_{atm}}\geq 0.8$. In most of the Superkamiokande allowed region the bound by Chooz
\cite{Chooz} essentially excludes $\nu_e\rightarrow\nu_{\mu}$ oscillations for atmospheric neutrino
oscillations. Furthermore the last results from Superkamiokande allow a solution of the
solar neutrino deficit in terms of
$\nu_e$ disappearance vacuum oscillations (as opposed to MSW \cite{MSW} oscillations within the sun)
with $\Delta m^2_{sol}\sim ~10^{-10}~eV^2$ and again nearly maximal mixing angles. Among the
large and small angle MSW solutions the small angle one is perhaps more likely  at the moment
(with \cite{Bahcall} $\Delta m^2_{sol}\sim 0.5~10^{-5}~eV^2$ and $\sin^2{2\theta_{sol}}\sim
5.5~10^{-3}$) than the large angle MSW solution. Of course experimental uncertainties are
still large and the numbers given here are presumably only indicative. But by now it is very unlikely that all this
evidence for neutrino oscillations will disappear or be explained away by astrophysics or other solutions. The
consequence is that we have a substantial evidence that neutrinos are massive.

In a strict minimal standard model point of view neutrino masses could vanish if no right handed neutrinos
existed (no Dirac mass) and lepton number was conserved (no Majorana mass). In Grand Unified theories both these
assumptions are violated. The right handed neutrino is required in all unifying groups larger than SU(5). In SO(10)
the 16 fermion fields in each family, including the right handed neutrino, exactly fit into the 16 dimensional
representation of this group. This is really telling us that there is something in SO(10)! The SU(5)
alternative in terms of $\bar 5+10$, without a right handed neutrino, is certainly less elegant. The breaking of
$|B-L|$, B and L is also a generic feature of Grand Unification. In fact, the see-saw mechanism \cite{ssm} explains
the smallness of neutrino masses in terms of the large mass scale where $|B-L|$ and L are violated. Thus, neutrino
masses, as would be proton decay, are important as a probe into the physics at the GUT scale.

Oscillations only determine squared mass differences and not masses. The case of three nearly degenerate neutrinos
is the only one that could in principle accomodate neutrinos as hot dark matter together with solar and atmospheric
neutrino oscillations. According to our previous discussion, the common mass should be around 1-3 eV. The solar
frequency could be given by a small 1-2 splitting, while the atmospheric frequency could be given by a still small
but much larger 1,2-3 splitting. A strong constraint arises in the degenerate case from neutrinoless double beta
decay which requires that the ee entry of
$m_{\nu}$ must obey
$|(m_{\nu})_{11}|\leq 0.46~{\rm eV}$. As observed in ref. \cite{GG}, this bound can only be 
satisfied if
double maximal mixing is realized, i.e. if also solar neutrino oscillations occur with nearly maximal mixing.
We have mentioned that it is not at all clear at the moment that a hot dark matter component is really
needed \cite{kol}. However the only reason to consider the fully degenerate solution is 
that it is compatible
with hot dark matter.
Note that for degenerate masses with $m\sim 1-3~{\rm eV}$ we need a relative splitting $\Delta m/m\sim
\Delta m^2_{atm}/2m^2\sim 10^{-3}-10^{-4}$ and an even smaller one for solar neutrinos. It is difficult
to imagine a natural mechanism compatible with unification and the see-saw mechanism to arrange such a
precise near symmetry.

If neutrino masses are smaller than for cosmological relevance, we can have the hierarchies $|m_3| >> |m_{2,1}|$
or $|m_1|\sim |m_2| >> |m_3|$. Note that we
are assuming only two frequencies, given by $\Delta_{sun}\propto m^2_2-m^2_1$ and
$\Delta_{atm}\propto m^2_3-m^2_{1,2}$. We prefer the first case, because for quarks and leptons one
mass eigenvalue, the third generation one, is largely dominant. Thus the dominance of $m_3$ for neutrinos
corresponds to what we observe for the other fermions.  In this case, $m_3$ is determined by the atmospheric
neutrino oscillation frequency to be around $m_3\sim0.05~eV$. By the see-saw mechanism $m_3$ is related to some
large mass M, by $m_3\sim m^2/M$. If we identify m with either the Higgs vacuum expectation value or the top mass
(which are of the same order), as suggested for third generation neutrinos by Grand Unification in simple SO(10)
models, then M turns out to be around $M\sim 10^{15}~GeV$, which is consistent with the connection with GUT's. If
solar neutrino oscillations are determined by vacuum oscillations, then $m_2\sim 10^{-5}~eV$ and we have that the
ratio $m_2/m_3$ is well consistent with $(m_c/m_t)^2$.

A lot of attention is being devoted to the
problem of a natural explanation of the observed nearly maximal mixing angle for atmospheric
neutrino oscillations and possibly also for solar neutrino oscillations, if explained by vacuum
oscillations\cite{AF}. Large mixing angles are somewhat unexpected because
the observed quark mixings are small and the quark, charged lepton and neutrino mass matrices are to
some extent related in GUT's. There must be some special interplay between the neutrino Dirac
and Majorana matrices in the see-saw mechanism in order to generate maximal
mixing. It is hoped that looking for a natural explanation of large neutrino mixings can lead us to decripting
some interesting message on the physics at the GUT scale.

\section{Comparing the Data with the Minimal Supersymmetric Standard Model}

The MSSM \cite{43} is a completely specified,
consistent and computable theory. There are too many parameters to attempt a direct fit of the data to
the most general framework. So we consider two significant limiting cases: the "heavy" and the
"light" MSSM.

	The "heavy" limit corresponds to all s-particles being sufficiently massive, still within the limits
of a natural explanation of the weak scale of mass. In this limit a very important result holds
\cite{58}: for what concerns the precision electroweak tests, the MSSM predictions tend to reproduce
the results of the SM with a light Higgs, say $m_H\sim$ 100 GeV. So if the masses of SUSY partners are pushed
at sufficiently large values the same quality of fit as for the SM is guaranteed. Note that for $m_t\sim175~GeV$
and $m_H\sim70~GeV$ the values of the four epsilons computed in the SM lead to a fit of the corresponding
experimental values with
$\chi^2\sim4$, which is reasonable for $d.o.f=4$. This value corresponds to the fact that the central values of
$\epsilon_1$,$\epsilon_2$, $\epsilon_3$ and -$\epsilon_b$ are all below the SM value by about $1\sigma$, as can
be seen from fig.11.

	In the "light" MSSM option some of the superpartners have a relatively small mass, close to their
experimental lower bounds. In this case the pattern of radiative corrections may sizeably deviate from
that of the SM \cite{pok}. The potentially largest effects occur in vacuum polarisation amplitudes and/or the
$Z\rightarrow b\bar b$  vertex. In particular we recall the following contributions :

	i) a threshold effect in the Z wave function renormalisation \cite{58} mostly due to the vector
coupling of charginos and (off-diagonal) neutralinos to the Z itself. Defining the vacuum polarisation
functions by $\Pi_{\mu\nu}(q^2)=-ig_{\mu\nu}[A(0)+q^2 F(q^2)]+q_\mu q_\nu$ terms, this is a positive
contribution to $\epsilon_5=m^2_Z  F'_{ZZ} (m^2_Z)$,the prime denoting a derivative with respect to
$q^2$ (i.e. a contribution to a higher derivative term not included in the usual epsilon formalism). The
$\epsilon_5$ correction shifts $\epsilon_1$, $\epsilon_2$ and $\epsilon_3$ by -$\epsilon_5$,
-$c^2\epsilon_5$ and  -$c^2\epsilon_5$ respectively, where $c^2=\cos^2{\theta_W}$, so that all of them
are reduced by a comparable amount. Correspondingly all the Z widths are reduced without affecting the
asymmetries. This effect falls down particularly fast when the lightest chargino mass increases from a
value close to $m_Z$/2. Now that we know,  from the LEP2 runs, that the chargino mass is
 not smaller than $m_Z$ its possible impact is drastically reduced.
 
	ii) a positive contribution to $\epsilon_1$ from the virtual exchange of split multiplets of SUSY partners,
for example of the scalar top and bottom  superpartners \cite{59}, analogous to the contribution of the
top-bottom left-handed quark doublet. From the experimental value of $m_t$ not
much space is left for this possibility, and the experimental value of $\epsilon_1$ is an important constraint
on the spectrum. This is especially true now that the rather large lower limits on the chargino mass reduce the
size of a possible compensation from $\epsilon_5$ .For example, if the stop is light then it must be mainly a
right-handed stop. Also large values of $\tan\beta$ are disfavoured because they tend to enhance the splittings
among SUSY partner multiplets. In general it is simpler to decrease the predicted values of $\epsilon_2$ and
$\epsilon_3$ by taking advantage of $\epsilon_5$ than to decrease $\epsilon_1$, because the negative shift
from $\epsilon_5$ is most often counterbalanced by the increase from the effect of split SUSY multiplets.

	iii) a negative contribution to $\epsilon_b$ due to the virtual exchange of a charged Higgs
\cite{60}. If one defines, as customary, $\tan{\beta}=v_2/v_1$ ($v_1$ and $v_2$ being the vacuum
expectation values of the Higgs doublets giving masses to the down and up quarks, respectively), then,
for negligible bottom Yukawa coupling or $\tan{\beta}<< m_t/m_b$, this contribution is proportional to
$m^2_t$ /$\tan^2{\beta}$.

	iv) a positive contribution to $\epsilon_b$ due to virtual chargino--s-top exchange \cite{61} which
in this case is proportional to $m^2_t$ /$\sin^2{\beta}$ and prefers small $\tan\beta$. This effect
again requires the chargino and the  s-top to be light in order to be sizeable.

With the recent limits set by LEP2 on the masses of SUSY partners the above effects are small enough that other
contributions from vertex diagrams could be comparable. Thus in the following we will only consider the
experimental values of the epsilons obtained at the level denoted by "All Asymmetries" which only assumes lepton
universality.

We have analysed the problem of what configurations of masses in the "light" MSSM are favoured or disfavoured
by the present data (\cite{ABC98},updating ref.{\cite{63}). We find that no lower limits on the masses of SUSY partners are
obtained which are better than the direct limits. One exception is the case of s-top and s-bottom  masses, which
are severely constrained by the $\epsilon_1$ value and also, at small $\tan\beta$, by the increase at LEP2 of the
direct limit on the Higgs mass. Charged higgs masses are also rather severely constrained. Since the central
values of
$\epsilon_1$,$\epsilon_2$ and 
$\epsilon_3$ are all below the SM it is convenient to make $\epsilon_5$ as large as possible. For this purpose
light gaugino and s-lepton masses are favoured. We find that for $m_{\chi^+_1}\sim 90-120~GeV$ the effect is
still sizeable. Also favoured are small values of $\tan\beta$ that allow to put s-lepton masses relatively low,
say, in the range 100-500~GeV, without making the split in the isospin doublets too large for $\epsilon_1$.
Charged Higgses must be heavy because they contribute to $\epsilon_b$ with the wrong sign.  A light right-handed
s-top could help on $R_b$ for a higgsino-like chargino. But one needs small mixing (the right-handed s-top must be
close to the mass eigenstate) and beware of the higgs mass constraint at small $\tan\beta$ (a higgs mass above
83~GeV, the range of LEP2 at $\sqrt{s}=183~GeV$, starts being a strong constraint at small $\tan\beta$). So we
prefer in the following to keep the s-top mass large. The limits on $b\rightarrow s\gamma$ also prefer heavy charged higgs
and s-top \cite{70}.

\section{The LEP2 Programme and the Search for the Higgs and New Physics}

The LEP2 programme has started in the second part of 1996. At first the energy has been fixed at 161~GeV, which is the
most favourable energy for the measurement of $m_W$ from the cross-section for
$e^+e^- \rightarrow W^+W^-$ at threshold. Then gradually the energy was brought up to 172, 183, 189 GeV. It will be
increased up to a maximum of about 200~GeV to be reached in mid '99. An integrated luminosity of about
150~pb$^{-1}$  per year is now achievable (in fact more was achieved in 1998). LEP2 has been approved to run until the end of 2000,
before the shutdown for the installation of the LHC. The main goals of LEP2 are the search for the Higgs and for new
particles, the measurement of
$m_W$ and the investigation of the triple gauge vertices
$WWZ$ and $WW\gamma$.  A complete updated survey of the LEP2 physics is collected in the two volumes of ref.~\cite{lep2}. 

	An important competitor of LEP2 is the Tevatron collider. In mid 2000 the Tevatron will start RunII with the purpose
of collecting a few $fb^{-1}$ of integrated luminosity at $2~TeV$. The competition is especially on the search of
new particles, but also on
$m_W$ and the triple gauge vertices. For example, for supersymmetry while the Tevatron is superior for gluinos and
squarks,  LEP2 is strong on Higgses, charginos, neutralinos and sleptons. There are plans for RunIII to start in 2002
or so with the purpose
of collecting of the order  $5~fb^{-1}$ of integrated luminosity per year. Then the Tevatron could also hope to find
the Higgs before the LHC starts if the Higgs mass is close to the LEP2 range.

	Concerning the Higgs it is interesting to recall that the large value of $m_t$ has important implications on
$m_H$ both in the minimal SM \cite{zziii}$-$\cite{bbiiii} and in its minimal supersymmetric
extension\cite{cciiii}$,$\cite{ddiiii}. I will now discuss the restrictions on $m_H$ that follow from the observed value
of
$m_t$.

	It is well known\cite{zziii}$-$\cite{bbiiii} that in the SM with only one Higgs doublet a lower limit on
$m_H$ can be derived from the requirement of vacuum stability. The limit is a function of $m_t$ and of the energy scale
$\Lambda$ where the model breaks down and new physics appears. Similarly an upper bound on $m_H$ (with mild dependence
on $m_t$) is obtained \cite{eeiiii} from the requirement that up to the scale $\Lambda$ no Landau pole appears. The
lower limit on
$m_H$ is particularly important in view of the search for the Higgs at LEP2. Indeed the issue is whether one can reach
the conclusion that if a Higgs is found at LEP2, i.e. with $m_H \leq m_Z$, then the SM must break down at some scale
$\Lambda  >$ 1~TeV. 

	The possible instability of the Higgs potential $V[\phi]$ is generated by the quantum loop corrections to the
classical expression of $V[\phi]$. At large $\phi$ the derivative $V'[\phi]$ could become negative and the potential
would become unbound from below. The one-loop corrections to $V[\phi]$ in the SM are well known and change the dominant
term at large $\phi$ according to $\lambda \phi^4 \rightarrow (\lambda +
\gamma~{\rm log}~\phi^2/\Lambda^2)\phi^4$. The one-loop approximation is not enough for our purposes, because it fails
at large enough $\phi$, when
$\gamma~{\rm log}~\phi^2/\Lambda^2$ becomes of order 1. The renormalization group improved version of the corrected
potential leads to the replacement $\lambda\phi^4 \rightarrow
\lambda(\Lambda)\phi'^4(\Lambda)$ where $\lambda(\Lambda)$ is the running coupling and
$\phi'(\mu) = {\rm exp}\int^t \gamma(t')dt'\phi$, with $\gamma(t)$ being an anomalous dimension function and $t = {\rm
log}\Lambda/v$ ($v$ is the vacuum expectation value
$v = (2\sqrt 2 G_F)^{-1/2}$). As a result, the positivity condition for the potential amounts to the requirement that
the running coupling $\lambda(\Lambda)$ never becomes negative. A more precise calculation, which also takes into
account the quadratic term in the potential, confirms that the requirements of positive
$\lambda(\Lambda)$ leads to the correct bound down to scales $\Lambda$ as low as $\sim$~1~TeV. The running of
$\lambda(\Lambda)$ at one loop is given by: 
\begin{equation}
\frac{d\lambda}{dt} = \frac{3}{4\pi^2} [ \lambda^2 + 3\lambda h^2_t - 9h^4_t + {\rm gauge~terms}]~,
\label{131}
\end{equation} with the normalization such that at $t=0, \lambda = \lambda_0 = m^2_H/2v^2$ and the top Yukawa coupling
$h_t^0 = m_t/v$. We see that, for $m_H$ small and $m_t$ large,
$\lambda$ decreases with $t$ and can become negative.  If one requires that
$\lambda$ remains positive up to $\Lambda = 10^{15}$--$10^{19}$~GeV, then the resulting bound on $m_H$ in the SM with
only one Higgs doublet is given by \cite{aaiiii}:
\begin{equation} m_H > 134 + 2.1 \left[ m_t - 173.8 \right] - 4.5~\frac{\alpha_s(m_Z) - 0.119}{0.006}~.
\label{25h}
\end{equation}

Summarizing, we see that the discovery of a Higgs particle at
LEP2, or $m_H\lappeq 100~GeV$, would imply that the SM breaks down at a scale
$\Lambda$ of the order of a few TeV. It can be shown \cite{zziii} that the lower limit is not much relaxed even if strict
vacuum stability is replaced by some sufficiently long metastability.

The upper limit on
the Higgs mass in the SM is important for assessing the chances of success of the LHC as an accelerator designed
to solve the Higgs problem. The upper limit \cite{eeiiii} arises from the requirement that the Landau pole associated with the non
asymptotically free behaviour of the $\lambda \phi^4$ theory does not occur below the scale $\Lambda$. The initial value of
$\lambda$ at the weak scale increases with $m_H$ and the derivative is positive at large $m_H$. Thus if $m_H$ is too large the
Landau pole occurs at too low an energy. The upper limit on $m_H$ has been recently reevaluated
\cite{hr}. For
$m_t\sim 175~GeV$ one finds
$m_H\lappeq 180~GeV$ for $\Lambda\sim M_{GUT}-M_{Pl}$ and $m_H\lappeq 0.5-0.8~TeV$ for $\Lambda\sim
1~TeV$. Actually, for
$m_t \sim$ 174~GeV, only a small range of values for $m_H$ is allowed, $130 < m_H <~\sim 200$~GeV, if the SM holds up
to $\Lambda \sim M_{GUT}$ or $M_{Pl}$. 

 A particularly
important example of theory where the above bounds do not apply and in particular the lower bound is violated is the MSSM,
which we now discuss. As is well known \cite{43}, in the MSSM there are two Higgs doublets, which implies three neutral
physical Higgs particles and a pair of charged Higgses. The lightest neutral Higgs, called $h$, should be lighter than
$m_Z$ at tree-level approximation. However, radiative corrections \cite{ffiiii} increase the $h$ mass by a term
proportional to $m^4_t$ and  logarithmically dependent on the stop mass . Once the radiative corrections are taken into
account the $h$ mass still remains rather small: for $m_t$ = 174~GeV one finds the limit (for all values of tg
$\beta)~m_h < 130$~GeV
\cite{ddiiii}. Actually there are reasons to expect that $m_h$ is well below the bound. In fact, if $h_t$ is large at
the GUT scale, which is suggested by the large observed value ot $m_t$ and by a natural onsetting of the electroweak
symmetry breaking induced by $m_t$, then at low energy a fixed point is reached in the evolution of $m_t$. The fixed
point corresponds to $m_t \sim 205 \sin\beta$~GeV (a good approximate relation for tg $\beta = v_{up}/v_{down} < 10$).
If the fixed point situation is realized, then $m_h$ is considerably below the bound, as shown in ref.\cite{ddiiii}.

In conclusion, for $m_t \sim 174$~GeV, we have seen that, on the one hand, if a Higgs is found at LEP the SM cannot be
valid up to $M_{Pl}$. On the other hand, if a Higgs is found at LEP, then the MSSM has good chances, because this model
would be excluded for $m_h > 130$~GeV.

	For the SM Higgs, which plays the role of a benchmark also important for a more general context, the LEP2 reach has
been studied in detail \cite{lep2}. At $200~GeV$ with about $150~pb^{-1}$ per experiment one can discover or exclude a SM Higgs
up to about $105~GeV$ of mass. In the MSSM a more complicated discussion is needed because there are several Higgses and the
parameter space is multidimensional. Only the lightest MSSM Higgs h is accessible at LEP2. The dominant production
processes are $e^+e^-->hZ$ and $e^+e^-->hA$, where A is the CP odd MSSM Higgs particle. They are nicely complementary. At
given $m_h$ within the range of interest, at large $\tan{\beta}$ the first process is the relevant one, while the second
determines the bound at small $\tan{\beta}$.   The absolute lower limit on
$m_h$ for a given beam energy and integrated luminosity is always below the limit on the SM Higgs, because the crossections
are smaller. For example the present limit is around $90~GeV$ for the SM Higgs and around $80~GeV$ for the MSSM Higgs. It is
interesting that by the end of LEP2 one will have completely explored the region at small $\tan{\beta}$ (below a value of
about 5), which is a particularly likely region.

A main goal of LEP2 is the search for supersymmetry. For charginos the discovery range at LEP2 is only limited by the beam
energy for practically all values of the parameters. In fact the typical limit is at present about $90~GeV$. Thus every
increase of the beam energy is directly translated into the upper limit in chargino mass for discovery or exclusion. For the
Tevatron the discovery range is much more dependent on the position in parameter space. For some limited regions of this
space, with
$1~fb^{-1}$ of integrated luminosity, the discovery range for charginos at the Tevatron goes well beyond
$m_\chi = 100~GeV$, i.e. the boundary of LEP2, but in much of the parameter space LEP2 at the maximum energy would
be sensitive to larger chargino masses.. 

	The stop is probably the lightest squark. For a light stop the most likely decay modes are  $\tilde t
\rightarrow b\chi^+$ if kinematically allowed, otherwise  
$\tilde t \rightarrow c \chi$. At LEP2 the discovery range is up to about $(E_{beam}-10)$~GeV. At
the Tevatron there is some difference between the two possible decay modes and some dependence on the position in the 
$\tilde t - \chi$ or the $\tilde t - \chi^+$ planes, but in general the Tevatron is very powerful for s-quarks and gluinos
and much of the LEP2 range is already excluded by the Tevatron.

By now most of the discovery potential of LEP2 for supersymmetry has been already deployed. For example, the limit on the
chargino mass was about $45~GeV$ after LEP1 and is now about $90~GeV$ and can only improve up to $100~GeV$. For the Higgs
the experimental task is more demanding and so one is only a bit more than half way through: the lower limit on the SM Higgs
was around
$67~GeV$ after LEP1, is now about $90~GeV$ and could go up to $105~GeV$ or so. So there are still fair chances for LEP2 to
find the Higgs, especially because the attainable range of masses is particularly likely in the MSSM.

The study of the $W^+W^-$ crosssection is a very important chapter of LEP2 physics \cite{lep2}. In the Born approximation three
Feynman diagrams contribute to the crosssection, as shown in fig.12. In the two s-channel exchange diagrams the triple gauge
vertices
$WW\gamma$ and $WWZ$ appear, while the third is the t-channel neutrino exchange that only involves well established charged
current couplings. One loop radiative corrections have also been computed . It is interesting that if we take neutrino exchange
alone or neutrino plus
$\gamma$ exchanges alone the crosssection increases much faster with energy than the complete result. This corresponds to the
good convergence properties of the SM which in fact is renormalisable. Indeed, the WW crosssection is related to the imaginary
part of the WW loop contribution to the amplitude for $e^+e^-\rightarrow e^+e^-$. The good large energy behavior of the
former crosssection is related to the convergence ot the latter loop correction. The data neatly confirm the SM prediction as
shown in fig.13. Thus the WW crosssection supports the specific form of the triple gauge vertices as predicted by the SM. More
detailed studies with large statistics are useful to set bounds on possible departures from the exact SM predicted
couplings. In fact the study of triple gauge vertices is another major task of LEP2. The capabilities of LEP2 in this domain
are comparable to those of the LHC. LEP2 can push down the existing direct limits considerably down. For
given anomalous couplings the departures from the SM are expected to increase with energy. For the energy and the
luminosity available at LEP2, given the accuracy of the SM established at LEP1, it is however not very likely, to find
signals of new physics in the triple gauge vertices.
%%%%%%%%%%%%%%%%%%%%%%%%%%%%%%%%%
\begin{figure}
\hglue 2.0cm
\epsfig{figure=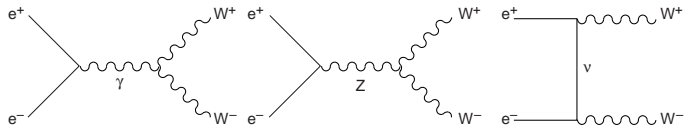, width=9cm}
\caption[]{Lowest order Feynman diagrams for $e^+e^- \rightarrow W^+W^-$.}
\end{figure}
%%%%%%%%%%%%%%%%%%%%%%%%%%%%%%%%%%
%%%%%%%%%%%%%%%%%%%%%%%%%%%%%%%%%
\begin{figure}
\hglue 2.0cm
\epsfig{figure=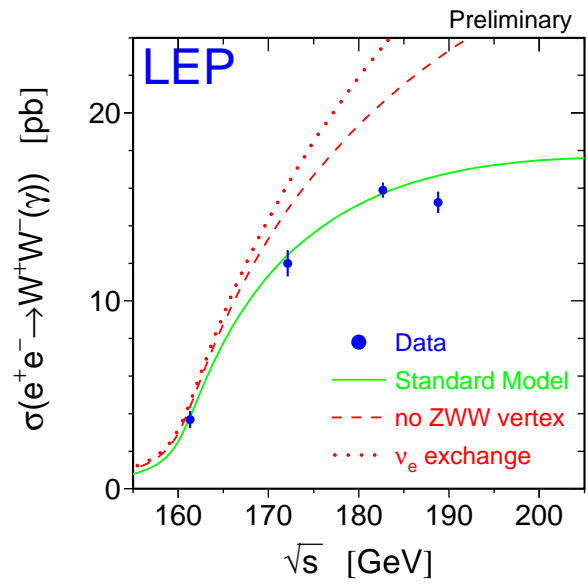, width=8.0cm}
\caption[]{Data vs theory for the WW cross-section measured at LEP2.  The solid line is the SM prediction.  The dashed
and dotted lines refer to only a subset diagrams as indicated.}
\end{figure}
%%%%%%%%%%%%%%%%%%%%%%%%%%%%%%%%%%
The measurement of $m_W$ is been done at LEP2 from the $W^+W^-$ cross-section at threshold and from direct reconstruction of
the W mass from the final state after W decay. At present
$m_W$ is known with an error of $\pm 60$~MeV from the combined LEP2 and Tevatron direct measurements (see table~1), with the
same error of $\pm90~MeV$ at LEP2 and at the Tevatron. From the fit to all electroweak data one finds $m_W = 80370 \pm 27~MeV$
(see eq.(\ref{10car})), in agreement with the direct measurement. As a consequence the goal for LEP2 is to measure $m_W$ with
an accuracy $\delta m_W
\leq \pm (30-40)$~MeV, in order to provide an additional significant check of the theory. 

For the threshold method \cite{lep2} the minimum of the statistical error is obtained for
$\sqrt s = 2m_W + 0.5$~GeV = 161 GeV, which in fact was the initial operating energy of LEP2. At threshold the WW
crossesection is dominated by the neutrino t-channel exchange which is quite model independent. The total error of this method
is dominated by the statistics. With the collected luminosity at 161 GeV of
$\sim 10~pb^{-1}$ per experiment, the present combined result is
$m_W = (80.4\pm 0.2 \pm 0.03)$~GeV
\cite{ew}. Thus with the available data at threshold this method is not sufficient by itself.

	In principle the direct reconstruction method can use the totally hadronic or the semileptonic final states
$e^+e^- \rightarrow W^+W^- \rightarrow jjjj$ or $jjl\nu$. The total branching ratio of the hadronic modes is 49\%,
while that of the $\ell = e,\mu$ semileptonic channels is 28\%. The hadronic channel has more statistics but could be
severely affected by non-perturbative strong interaction effects: colour recombination among the jets from different
W's and Bose correlations among mesons in the final state from WW overlap. Colour recombination is perturbatively
small. But gluons with $E < \Gamma_W$ are important and non-perturbative effects could be relatively large, of the
order of 10--100~MeV. Similarly for Bose correlations. One is not in a position to really quantify the associated
uncertainties. Fortunately the direct reconstruction from the semi-leptonic channels can, by itself, lead to a total
error $\delta m_W = \pm 44~MeV$,  for the combined four experiments, each with 500~pb$^{-1}$ of luminosity collected at
$\sqrt s \geq 175$~GeV. Thus the goal of measuring $m_W$ with an accuracy below  $\delta m_W = \pm 50$~MeV can be
fulfilled, and it is possible to do better by learning from the data how to limit the error from colour recombination
and Bose correlations.

\section{Conclusion}

Today in particle physics we follow a double approach: from above and from below. From above there are, on the theory
side, quantum gravity (that is superstrings), GUT theories and cosmological scenarios. On the experimental side there
are underground experiments (e.g. searches for neutrino oscillations and proton decay), cosmic ray
observations, satellite experiments (like COBE, IRAS etc) and so on. From below, the main objectives of theory and
experiment are the search of the Higgs and of signals of particles beyond the Standard Model (typically supersymmetric
particles). Another important direction of research is aimed at the exploration of the flavour problem: study of CP
violation and rare decays. The general expectation is that new physics is close by and that should be found very
soon if not for the complexity of the necessary experimental technology that makes the involved time scale painfully
long.

\end{document}